\documentclass{aastex63}
\usepackage{epstopdf}
\usepackage{natbib}
\usepackage{graphicx}
\usepackage{CJK}
\usepackage{booktabs}
\usepackage{tabularx}
\usepackage{multirow}
\usepackage[shortcuts]{extdash}
\usepackage{threeparttable}

\newcommand       \Teff         {T_{\rm {eff}}}

\newcommand       \BpRp[1]      {E_{\rm G_{BP},G_{RP}}^{#1}}
\newcommand       \BpminRp      {G_{\rm BP}-G_{\rm RP}}
\newcommand       \cBpRp[1]     {C_{\rm G_{BP},G_{RP}}^{#1}}
\newcommand       \JK[1]        {E_{\rm J,K_{s}}^{#1}}
\newcommand       \cJK[1]       {C_{\rm J,K_{s}}^{#1}}
\newcommand       \JminK        {J-K_{\rm s}}
\newcommand       \feh          {\rm \left[Fe/H \right]}
\newcommand       \logg         {{\rm log}\ g}

\submitjournal{ApJ}

\begin{document}
\begin{CJK*}{UTF8}{gbsn}

\title{Extinction of Taurus, Orion, Perseus and California Molecular Clouds Based on the LAMOST, 2MASS and Gaia surveys I: Three-dimensional Extinction and Structure}
\author[0000-0002-0316-1112]{ZheTai Cao (曹哲泰)}
\affiliation{Institute for Frontiers in Astronomy and Astrophysics,
            Beijing Normal University,  Beijing 102206, China}
\affiliation{Department of Astronomy,
               Beijing Normal University,
               Beijing 100875, China}
\author[0000-0003-3168-2617]{Biwei Jiang (姜碧沩)}
\affiliation{Institute for Frontiers in Astronomy and Astrophysics,
            Beijing Normal University,  Beijing 102206, China}
\affiliation{Department of Astronomy,
               Beijing Normal University,
               Beijing 100875, China}
\author[0000-0003-2645-6869]{He Zhao (赵赫)}
\affiliation{Purple Mountain Observatory,
               Chinese Academy of Sciences, Nanjing 210023, China}
\author[0000-0002-2473-9948]{Mingxu Sun (孙明旭)}
\affiliation{Department of Physics,
                Hebei Normal University, Shijiazhuang 050024, China}

\correspondingauthor{B.~W. Jiang}
\email{bjiang@bnu.edu.cn}

\begin{abstract}

The three-dimensional extinction and structure are studied for the Taurus, Orion, Perseus and California molecular clouds based on the LAMOST spectroscopy. Stellar color excess is calculated with the intrinsic color index derived from the atmospheric parameters in the LAMOST DR8 catalog and the observed color index in the Gaia EDR3 and the 2MASS PSC. In combination with the distance from the Gaia EDR3 parallax, the three-dimensional dust extinction maps are retrieved in the color excesses $E_{\rm{G_{BP},G_{RP}}}$ and $E_{\rm{J,K_{S}}}$ with an uncertainty of $\sim$0.03mag and $\sim$0.07mag respectively. The extinction maps successfully separate the clouds that overlap in the sky area and manifest the structure of the individual cloud. Meanwhile, a bow-like structure is found with a distance range from 175pc to 250pc, half of which is a part of the Per-Tau Shell in similar coordinates and distance while the other half is not. Three low-extinction rings are additionally discovered and briefly discussed.

\end{abstract}
\keywords{Distance measure (395); Interstellar dust (836); Molecular clouds (1072); Extinction (505); Interstellar dust extinction (837)}
\section{Introduction}\label{INTRO}

Molecular clouds (MCs), as the star birthplaces, are generally dense that causes high extinction. A precise estimation of the extinction is crucial to revealing the true brightness and color of the stars embedded and behind the cloud. In addition, molecular clouds are place for dust growth. The determination of the extinction law of MCs would help understand the dust evolution in various star-forming environments and the dust properties such as grain size. The nearby MCs to be studied in this work, specifically the Taurus MC (hereafter TMC), Orion MC (OMC), Perseus MC (PMC) and California MC (CMC), represent different including massive and low-mass star-forming environments. With precise measurements and large quantity of tracers, the extinction of MCs can be calculated with high precision and high spatial resolution, and therefore serve as the references of extinction for star forming regions.

Many works have been devoted to studying the distribution and properties of interstellar extinction, which cover star forming regions. The most widely used extinction map is the all-sky two dimensional reddening map by Schlegel, Finkbeiner and Davis \citep{sfd98} derived from dust infrared emission. With the distance measurements of billions of stars by Gaia \citep{GaiaDR2,GaiaEDR3_2021, GaiaDR3_summary}, some three-dimensional (3D) extinction maps are produced. For example, \citet{green15,Green19} created an extinction map that extends to a few kiloparsecs over three-quarters of the sky with the Pan\-/STARRS and 2MASS photometry. As for individual molecular cloud, \citet{TauCaliPer2010III, Orion2011IV} presented the near-infrared extinction map of several nearby molecular clouds that include TMC, PMC, CMC, OMC. \citet{OCTP4cloud} produced the continuous dust density and extinction maps of Orion, Cygnus X, Taurus, and Perseus MC by using the Gaussian Process. The Gaussian Process is also used to infer the structure of Orion A and California \citep{CaliOriA21_Kh,Orion2022_Kh}.

Most of the previous works used the photometric data to investigate the extinction of molecular clouds. Photometry has the advantage of being able to detect faint objects which leads to further distance and higher spatial resolution achievable. Spectroscopy, on the other hand, can determine the extinction more accurately with spectroscopically derived stellar parameters.  The development of multi-fiber observation has greatly increased the efficiency of spectroscopy. The H-band APOGEE survey takes 300 spectra at a time, which has accumulated almost six hundred thousand stellar spectra in DR17 \citep{APOGEE_review}. LAMOST, a reflective Schmidt telescope with a diameter of 5-m and a F.O.V of $5\degr$ that provides spectra for about 4000 objects in one exposure, has obtained stellar parameters for about ten millions stars \citep{Luo15_LMT}. Consequently, the calculation of high precision extinction becomes feasible for very large sample of stars from spectroscopy. Moreover, such calculation of color excess is independent of extinction law so that the multi-wavelength color-excess can be used to determine the extinction law to various molecular clouds and their dust properties.

This work intends to build the 3D extinction map and structure of the nearby MCs, specifically, Taurus, Orion, Perseus and California based on spectroscopic survey. Stellar intrinsic color indexes will be calculated from the atmospheric parameters derived from the LAMOST spectrum. In combination with the distance from \citet{Bjdis21} and photometery from 2MASS and Gaia EDR3\footnote{The recently released Gaia DR3 data contains the broad-band photometry already published as part of Gaia EDR3, and the astrometric data in Gaia DR3 are the same as those of Gaia EDR3 \citep{GaiaDR3_summary}.}, the extinction and the structure of the MCs are obtained. In brief, the color excess of each star is calculated in the Gaia bands (i.e. $E_{\rm{G_{BP},G_{RP}}}$) and in the 2MASS bands (i.e. $E_{\rm{J,K_{S}}}$), then a non-decreasing function is fitted to determine the variation of extinction with distance in the given sightline that reveals the 3D structure of the MC. Section \ref{SecData} describes the data from Gaia, 2MASS and LAMOST to be used. Section \ref{SecBE} and Section \ref{SecDis} present the methods to calculate the color excess and to decompose the MCs into distance slice. Section \ref{SecRes} discusses the 3D structures of the four molecular clouds one by one. Section \ref{Summary} is a summary.

\section{Data}\label{SecData}
\subsection{The Sample Selection}

The optical and near-infrared (NIR) bands are selected for measuring the clouds' extinction. Because the visual extinction is much larger than the near-infrared (the V band extinction is about ten times that in the K band \citep{Wang19_law}), the two wavelength ranges are expected to trace both the diffuse and dense clouds. Besides, the ratio of the optical-to-NIR extinction is an indicator of the extinction law to be studied in our next work.

The optical photometric data are taken from the space telescope Gaia for its high precision and full sky coverage. The adopted Gaia EDR3 data contains 1.5 billion sources whose photometry is accurate to mmag-level in the $G_{\rm BP}$, $G$, $G_{\rm RP}$ bands centering at 532, 673, and 797 nm respectively \citep{Jordi_2010_band}. The very wide band, $G$, is not used in this work because the extinction coefficient of this band varies strongly with stellar effective temperature and the extinction itself \citep{Dan18_G_extinction}. The NIR data are taken from the PSC of 2MASS that surveyed the whole sky in the $J$, $H$, $K_{\rm s}$ band, which bring about the photometry over more than 500 million sources with the average uncertainty better than 0.03mag \citep{2MASS_2006}.

The stellar parameters are taken from the LAMOST DR8 \citep{Luo15_LMT}. The LAMOST DR8 provides the effective temperature ($\Teff$), surface gravity (log $g$) and metallicity ($\feh$) for more than six-million stars. For the duplicated sources, only the parameters derived from the spectrum with the highest signal-to-noise ratio are kept. Then about 6.4 million stars are kept and cross-matched with the Gaia EDR3 and 2MASS PSC datasets respectively by a radius of $1\arcsec$. The distance is a key parameter to investigate the 3D extinction of the clouds. Here, we take the geometric distance provided by \citet{Bjdis21} which contains the distances and their uncertainties for 1.47 billion stars.

\begin{figure}
\centering
\centerline{\includegraphics[scale=1.0]{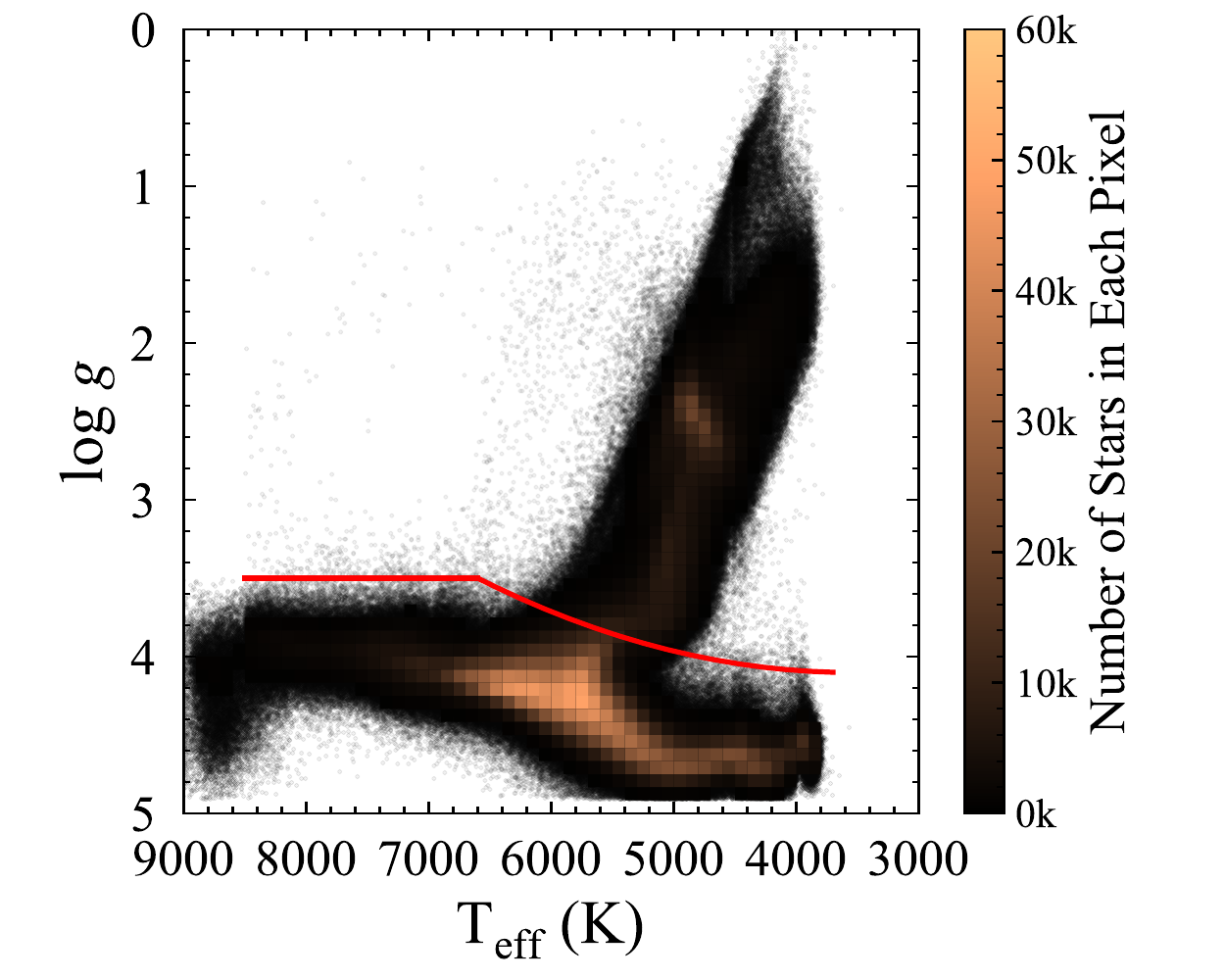}}
\caption{The Kiel diagram of the initial sample. The stars below the red line are selected as dwarfs, and the color indicates number density.
\label{SelectDwarfs}}
\end{figure}

The data quality is further restricted as following to exclude the poor measurements:
\begin{enumerate}
\item $\feh$ is within [-1.0, 0.5].
\item The error of $\Teff$, $\feh$ and ${\rm log}\ g$ is smaller than 150K, 0.15 dex and 0.3 dex respectively.
\item The signal-to-noise ratio in the $g$ band (the parameter ``snrg" in the LAMOST DR8 catalog) is larger than 10.
\item The photometric error  is $<0.05$mag in the Gaia EDR3 bands and $<0.1$mag in the 2MASS bands.
\end{enumerate}
With these limitations, $\sim$4.6 million are left for calculating the intrinsic color index  $\cBpRp{0}$ while $\sim$4.3 million stars are kept for $\cJK{0}$. Furthermore, only dwarf stars are retained for three reasons: (1) metallicity has less effect on the intrinsic colors of dwarfs than of giants  (see Figure 1 in \citealt{Zhao20}); (2) giants in the LAMOST catalog are mostly at a distance $>$ 1kpc, much further than the studied MCs; and (3) stellar parameters of dwarfs are more accurately determined from the LAMOST observations. In practice, the dwarf stars are selected from the Kiel diagram in Figure \ref{SelectDwarfs} by $\logg>3.8$ for $\Teff >6600K$ or $-0.0644 \times \Teff^2 + 0.457\times \Teff +3.09<\logg$ for $\Teff <6600K$. This reduces the sample to $\sim$ 3.2 million and 3.1 million stars for $\BpminRp$ and $\JminK$ respectively.

\subsection{The Cloud Regions}\label{cloudregion}
\citet{TauCaliPer2010III} defined the boundary of TMC in $(l,b) = ([165\degr,180\degr],[-10\degr,-20\degr]$). Later, \citet{TPshell21} recognized an ellipse substructure in the 3D space called the Tau Ring  centering at $(l,b,d) = (179\degr.5,\ -14\degr.2,\ 179{\rm pc})$ with a semimajor and a semiminor axis of 39pc and 26pc respectively, and a projection radius of about $10\degr$ at $d = 218$pc. Including all these structures,  we extend the boundaries a little to cover a larger area where the extinction is comparatively low,  and  select the boundary as $(l,b) = ([160\degr,195\degr],[10\degr,-40\degr]$.

The area selection for the other three MCs follows the same rule as for the TMC, i.e. we select an area that is extended a little from the previously defined. \citet{TauCaliPer2010III} delimited the PMC to $155^\circ<l<165^\circ$ and $-25^\circ<b<-15^\circ$, and \citet{CaliOriA21_Kh} defined $155^\circ<l<170^\circ$ and $-14^\circ<b<-6^\circ$ for CMC which agrees with \citet{TauCaliPer2010III}. For OMC that is composed of three major structures, \citet{Orion2011IV} suggested $203^\circ<l<217^\circ$ and $-21^\circ<b<-17^\circ$ for Orion A, $201^\circ<l<210^\circ$ and $-17^\circ<b<-5^\circ$ for Orion B and $188^\circ<l<201^\circ$ and $-18^\circ<b<-7^\circ$ for $\lambda$ Orionis. Taking all these into consideration, our selected regions of the four MCs are listed in the first and second column of Table \ref{Cloudboundary}. There are some overlaps in sightlines between some clouds such as TMC and CMC. Section \ref{SecDis} will introduce the method to further separate them in the 3D space.

\begin{figure}
\centering
\centerline{\includegraphics[scale=0.55]{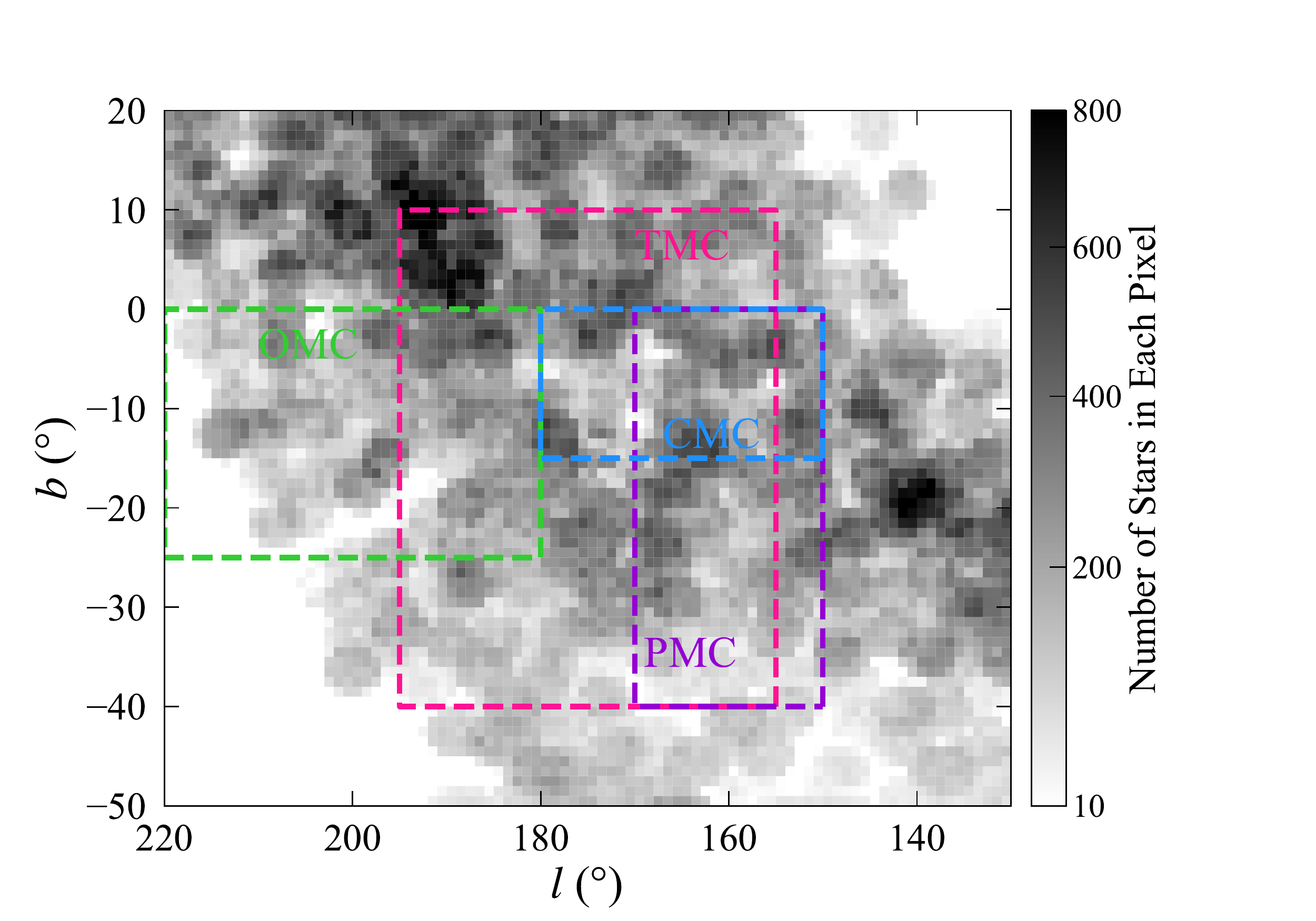}}
\caption{Number of stars in a circle with a radius of one degree in the studied region. The pixel with less than ten stars is marked by blank space.
\label{densityofstars}}
\end{figure}

By summing up the four clouds region in Table \ref{Cloudboundary},  the whole region to be studied is within $(l,b) = ([130\degr,220\degr],[-50\degr,+20\degr])$, a total area of about 5000 deg$^2$. The area of each cloud and the density of tracing stars are displayed in Figure \ref{densityofstars}.

\section{The intrinsic color index: the blue-edge method} \label{SecBE}

The blue-edge method takes the bluest observed color index as the intrinsic one for a given set of stellar parameters by assuming that the bluest star experiences little or no extinction among a large collective. It was first suggested by \citet{Jorge_2001_IRcolor} and then refined by \citet{Jian2017_RevBE} and \citet{W2014_IRlaw}, and applied in several works, e.g. in  \citet{Xue2016_IRlaw, Zhao18, Sun18_BE}. In practice, the relation of intrinsic color index with effective temperature is derived for a given luminosity class and a given range of metallicity since temperature is the primary factor to influence the color index. Although there have been some determinations of the relations in the above mentioned works, we re-construct the relation of the intrinsic color index, $\cBpRp{0}$ and $\cJK{0}$, with temperature for two reasons. One is that the $\BpminRp$ color is not included in previous works. The other is that the accuracy of the relation is improved by dividing the metallicity into more groups.

\begin{figure}
\centering
\centerline{\includegraphics[scale=0.63]{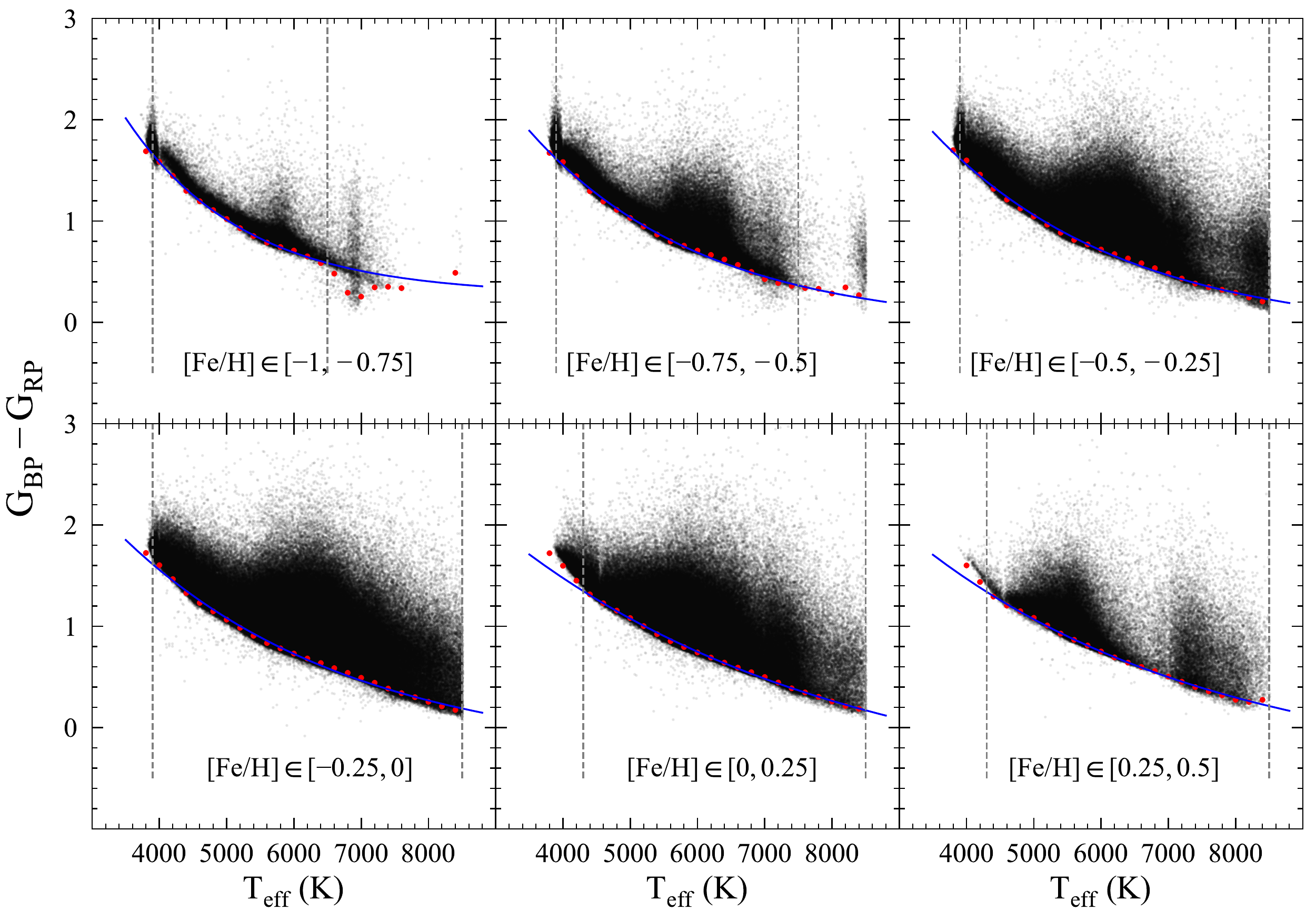}}
\caption{The $\Teff$ versus $G_{\rm BP}-G_{\rm RP}$ diagram of the sample dwarf stars in six metallicity bins. The black dots denote all the stars, while only stars between the two gray vertical lines are used to define the blue edge and calculate the intrinsic color index. The red dots denote the selected zero-reddening stars in each temperature bin and the blue line is the fitting curve with Equation~\ref{EqBE}.
\label{BEbprp}}
\end{figure}

The metallicity is divided into six groups with a step of 0.25 dex ranging from -1.0 to 0.5 dex to account for the effect on intrinsic color index by metallicity. Following \citet{Jian2017_RevBE}, we take the bluest 5$\%$ star in a given metallicity and temperature bin as the extinction-free star, and the color of the bluest 5$\%$ star is used to represent the intrinsic color for stars in the given bin that is shown by the red dots in Figure \ref{BEbprp}. Then, a curve is fit to derive the relation between the temperature and the color of the extinction-free stars in each metallicity bin. Consequently, the color excess is the difference of the observed and the intrinsic color index calculated from the curve. Here, the metallicity and temperature bins are 0.25\,dex and 200\,K respectively.

An exponential function with three free parameters is used to fit the extinction-free stars (i.e. the red dots in Figure \ref{BEbprp}):
\begin{equation}\label{EqBE}
C_{\lambda1,\lambda2}^0 = a\times \exp \left ( -\frac{\Teff}{b}  \right ) +c
\end{equation}

The case of $\BpminRp$ in six metallicity bins ranging from -1.00 to 0.50 dex is displayed in Figure \ref{BEbprp} while the case of $J-K_S$ is similar and not displayed . It can be seen that the range of $\Teff$ is cut at both the low and the high ends marked by vertical dashed lines where the stars are not numerous enough or the trend cannot be depicted by the same exponential function. The decrease of the sample due to this cutting would influence the results little, but guarantee the precision of the color index.

The uncertainty of the intrinsic color index comes from mainly photometric error and the blue-edge error. The mean photometric error and its standard deviation are $4\pm2$, $4\pm2$, $26\pm6$, and $32\pm17$mmag in the $G_{\rm BP}, G_{\rm RP}, J$, and $K_{\rm s}$  band respectively. The error induced by the bluest edge is about 30mmag for dwarfs \citep{Jian2017_RevBE}. In total, the uncertainty of $C_{\lambda1,\lambda2}^0$ is $\sim$30mmag for $\cBpRp{0}$ and $\sim $50mmag for $\cJK{0}$, which means that the major error comes from the blue-edge method for $\cBpRp{0}$  while from both photometry and the blue-edge method for  $\cJK{0}$.

The color excess is calculated straightforward by subtracting the intrinsic color index from the observed. Consequently, the error is about $\sim$30mmag for $\BpRp{}$  and $\sim$70mmag for $\JK{}$.
	

\section{Distance-sliced Extinction in the Area of the Molecular Clouds}\label{SecDis}

\subsection{The Extinction Map}\label{extinctionmap}

\begin{figure}
\centering
\centerline{\includegraphics[scale=0.5]{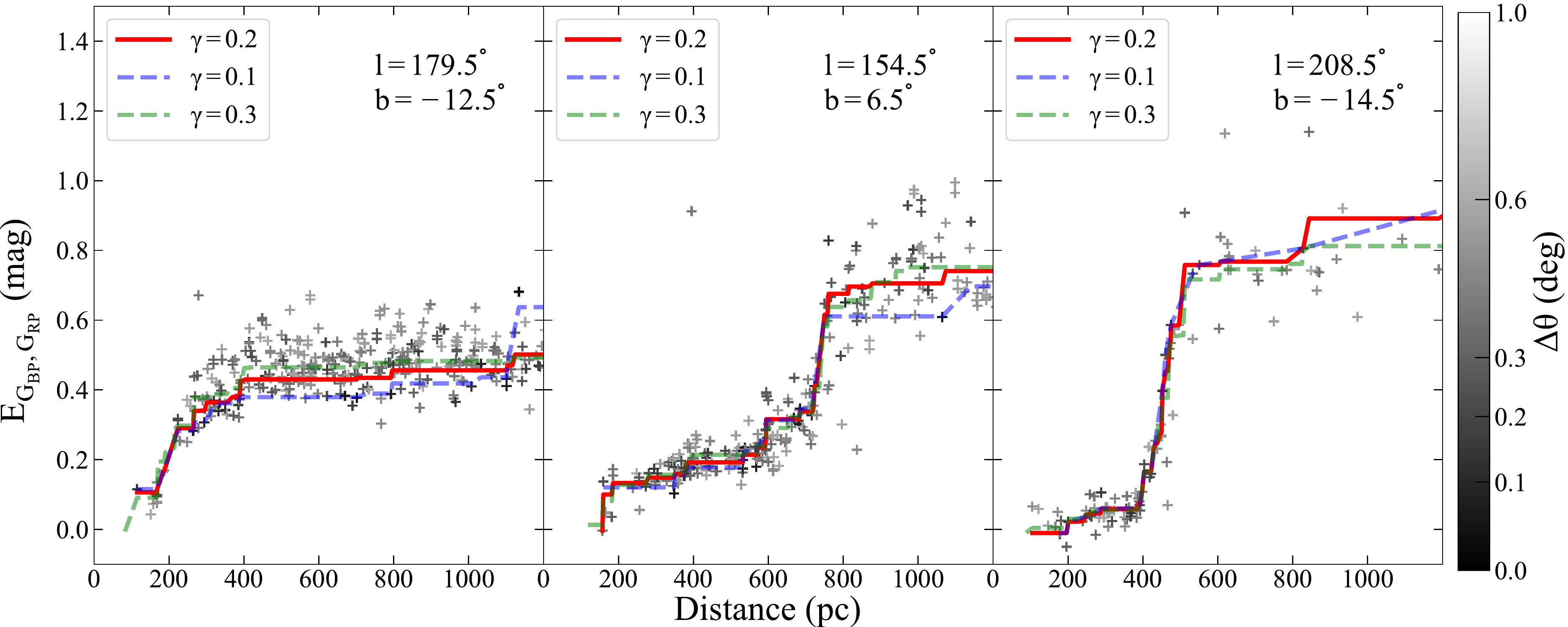}}
\caption{Test of $\gamma$ in Equation \ref{kernel} to derive the variation of the color excess $\BpRp{}$ with the distance towards three sightlines whose Galactic coordinates are indicated on the top right of each panel. The gray scale of the cross marks the angular distance to the sightline ($\Delta \theta$), and three dashed lines in different colors indicate the fitting result with different $\gamma$.
\label{sample}}
\end{figure}

The principle in deriving the 3D extinction in the area of the TMC, OMC, PMC and CMC is that the extinction along a sightline is a non-decreasing function of distance. With the distance by \citet{Bjdis21} and the color excess calculated above, a compromise between the number of tracers and the spatial resolution yields a step of 0.2\degr\ in both longitude and latitude. Due to the non-uniform extinction within a selected area of 0.2\degr\ squared and the error in distance and color excess, the variation of color excess with distance is scattering and sometimes not monotonically increasing as shown by the dots in Figure \ref{sample}.

The isotonic regression in the SCIKIT-LEARN package for PYTHON \citep{scikit-learn} is applied to trace the general tendency. For a given set of observations ($x_{1}$, $y_{1}$), ($x_{2}$, $y_{2}$),...,($x_{n}$, $y_{n}$), isotonic regression is a non-parametric regression that seeks a weighted least-square fit $\hat{y_{i}}$ for all $i$ in a monotonic model. It solves the following problem:
\begin{equation}\label{IR}
\min{\sum_{i=1}^{n}w_{i}(y_{i}-\hat{y_{i}})^{2}},\
{\rm subject\ to}\ {\hat {y}}_{i}\leq {\hat {y}}_{j}\ {\rm whenever}\ \displaystyle x_{i}\leq x_{j}
\end{equation}
where $y_{i}$ is the calculated color excess of star $i$, $\hat{y_{i}}$ is the fitted color excess along the given line of sight at the distance of star $i$, and $w_{i}$ is the weight. Typically, the weights are equal to 1 for all $i$. However, because the stars closer to the center of the given sightline are more representative, an exponential kernel function is adopted to describe $w_{i}$ as following:
\begin{equation}\label{kernel}
w_{i}= \left \{
\begin{array}{ll}
    \exp(-\frac{{\theta}^2}{2\gamma^{2}}),     &{\rm if}\ 0<\theta<\theta_{0} \\
    0,                              & {\rm otherwise}
\end{array}
\right.
\end{equation}
where $\theta$ is the angular distance of star $i$ to center of the selected sightline, $\gamma$ is a scale parameter, and $\theta_{0}$ defines the radius of the selected bin size. Meanwhile, the number of stars that have non-zero weights should be more than 10 in one sightline to ensure the credibility of the model fitting. Considering the existence of some sightlines where there are less than 10 objects within a 0.2\degr\ circle, a larger $\theta_{0}$ is selected. The scale parameter $\gamma$ is set to be equal to the resolution. Figure \ref{sample} shows the fitting results with three test values of $\gamma$ in three sightlines as examples. After testing the model with several values of resolution (i.e. $\gamma$),  $0.2^\circ$ and $1.0^\circ$ (5 times of $\gamma$) are selected respectively for $\gamma$ and $\theta_{0}$.

\begin{figure}
\centering
\centerline{\includegraphics[scale=0.7]{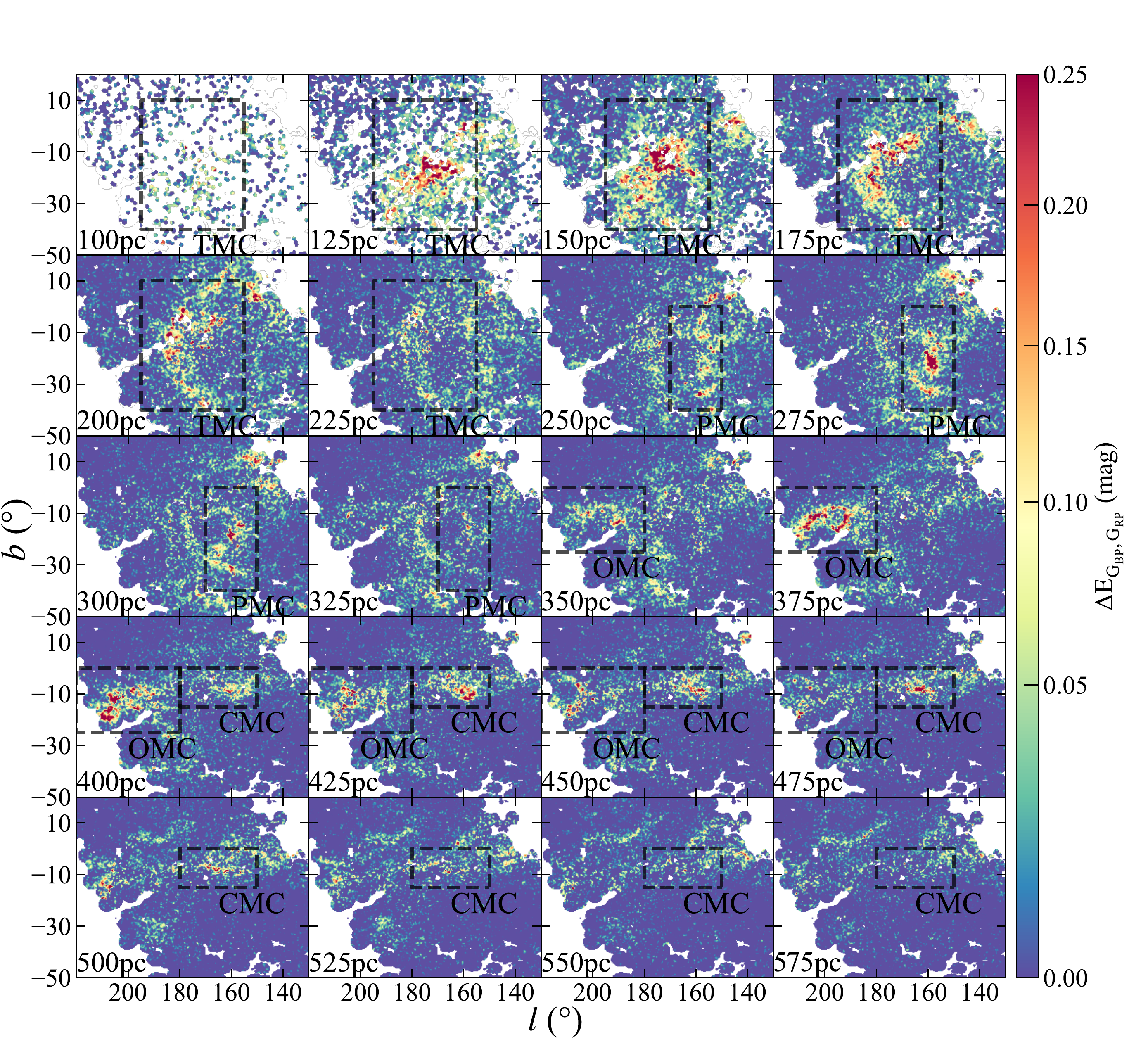}}
\caption{Extinctions in each distance slice at a step of 25pc, with the starting distance at the lower left of each graph.  The molecular cloud region is bordered by black box according to Table \ref{Cloudboundary}. The color excess in $\BpminRp$ in each distance bin is denoted by the color bar. The blank space marks the area with no reliable result. The sky area of each cloud is indicated by the dashed rectangle.
\label{slice}}
\end{figure}

The extinction is sliced every 25pc from 100pc to 600pc by the above non-parameter fitting, and the results are presented in Figure \ref{slice}. In combination with the continuity in the sky area, it can be seen that the four MCs appear in order of distance, i.e. TMC, PMC, OMC and CMC. The distance extents of each molecular cloud listed in Table \ref{Cloudboundary} are judged by observing the main extinction structure (high extinction areas) within its boundary in Figure \ref{slice}. It should be noted that the extension is partly caused by the substructures that appear at different distance instead of the thickness of the cloud. The details for each substructure will be discussed later.

\begin{figure}
\centering
\centerline{\includegraphics[scale=0.6]{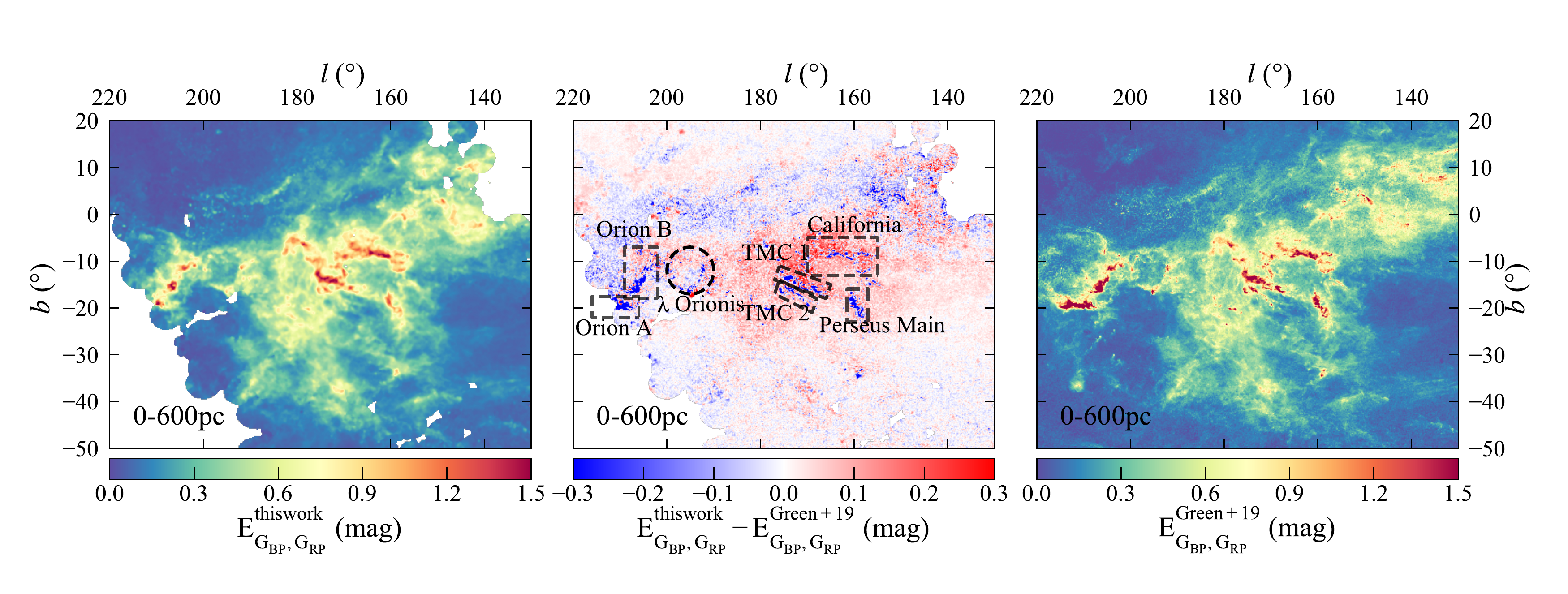}}
\caption{Left: the extinction map integrated to 600pc where the blank space means lack of data. Middle: the difference with \citet{Green19}, the blue color denotes the area where the extinction in \citet{Green19} is larger than ours and the red denotes that ours is larger. The prominent structures are marked by black dashed line. Right: the extinction map of \citet{Green19} integrated to a distance of 600pc.
\label{comparsionGreen}}
\end{figure}

\begin{figure}
\centering
\centerline{\includegraphics[scale=0.6]{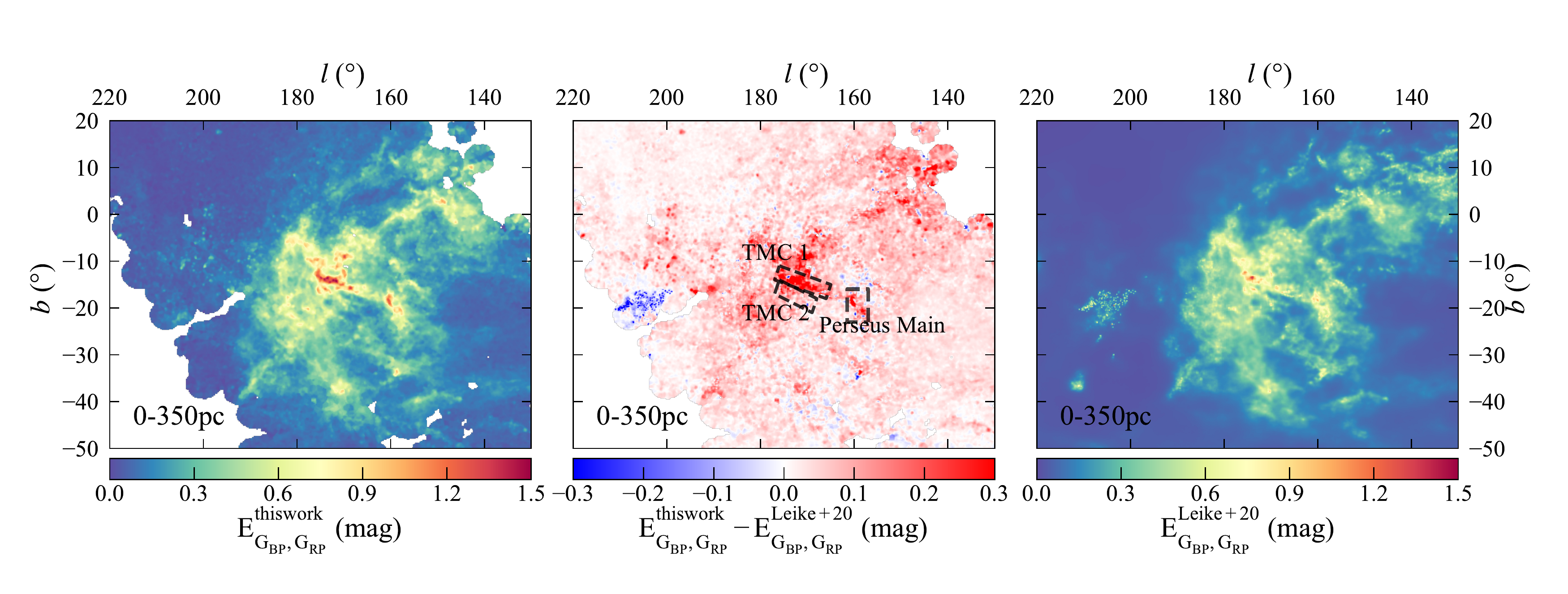}}
\caption{Comparison with \citet{L20}. The convention is the same as Figure \ref{comparsionGreen} except that the extinction is integrated out to a distance of 350pc.}
\label{comparsionLK}
\end{figure}

The results are compared  with \citet{Green19} and \citet{L20} in the same area and distance range. For comparing with \citet{Green19}, the integrated extinction map up to 600pc from this work and \citet{Green19} is displayed in Figure \ref{comparsionGreen}. The value of $E_{\rm{B,V}}^{\rm {Green}}$ in \citet{Green19} is converted to $\BpRp{}$ with the factor $E_{\rm{B,V}}^{\rm {Green}}/E_{\rm{G_{BP},G_{RP}}}=0.71\pm 0.01$ suggested by \citet{Sun21_G_BV}. It can be seen that the two results are roughly identical with a mean difference and its standard deviation of $0.007\pm0.093$mag respectively. Nevertheless, it is obvious that the extinction value is comparatively smaller in the high-extinction regions (the blue part in Figure \ref{comparsionGreen}), such as the dense MC regions marked by the black lines in the right panel in Figure \ref{comparsionGreen}. This can be explained by the relatively shallow depth of the LAMOST survey that has a limiting magnitude of about $15$ to $17$mag in the $g$ band, which is unable to detect the stars with a large extinction. The largest extinction in this work is about $2.0$mag in $\BpRp{}$ and $0.8$mag in $\JK{}$, equivalent to $A_{V}\sim 5$mag, occurring in some Galactic plane areas. This can be taken as the limiting depth of extinction in this work. For comparing with \citet{L20}, the integrated extinction map up to 350pc from this work and \citet{L20} is displayed in Figure \ref{comparsionLK}, where the factor that converts the $G$ band extinction to $\BpRp{}$ is 1.89 \citep{Wang19_law}. It can be seen that the results from this paper is slightly larger than that from \citet{L20} with a mean difference of $0.04\pm0.05$mag.

\subsection{The Distance to the Clouds}

Though the distance-sliced extinction gives the rough range of each cloud, the 25pc step is quite large. The distance to various parts of a molecular cloud can be more accurately determined. The basic method to determine the distance to the cloud is the same as that in \citet{Zhao20}  for supernova remnants (SNRs). It assumes that the extinction will present a sharp increase (i.e. a `jump') at the distance of the dense cloud due to its high dust grain density, and consequently the distance of the cloud is recognized in the variation of the interstellar extinction along the distance. Such model is used to analyze the distance to the extended sources (see e.g. \citealt{Chen17_s147} for MCs and \citealt{Zhao20} for SNRs). \citet{Zucker19_MBM,Zucker_20} used the change of reddening along with parallax to derive the distance to molecular clouds as well.

Consistent with the above analysis, a circle area with a radius of 1.0\degr\ is taken for the distance determination. For a given area, the extinction in terms of color excess is the function of distance:
\begin{equation}\label{CE_all}
E(d) = E^{\rm fgd}(d) + E^{\rm MC}(d)
\end{equation}
where $E(d)$ is the total color excess along the line of sight until $d$, $E^{\rm fgd}(d)$ is the extinction from the foreground diffuse medium, and $ E^{\rm MC}(d)$ is the contribution by the molecular cloud.  The $E^{\rm MC}(d)$ is described by a Gaussian error function:
\begin{equation}\label{CE_mc}
E^{\rm MC}(d) = \frac{\delta E}{2}\times \left [ 1+\mathit{erf}\left ( \frac{d-d^{\rm {MC}}}{\sqrt{2}\times \sigma} \right ) \right ]
\end{equation}
where $\delta E$, $d^{\rm {MC}}$, and $\sigma$ represent the extinction `jump', the distance to and the half thickness superposed onto the distance error of the specific cloud region in the given sightline respectively. However, instead of a two-order polynomial function used in \citet{Chen17_s147, Zhao20} or exponential function in \citet{Sun21_MCdistance}, a constant is assumed to describe the foreground extinction, i.e.:
\begin{equation}\label{CE_fg}
E^{\rm fgd}(d) = E_{0}
\end{equation}
This modification is to match the close distance of the clouds, which implies very small foreground extinction. Technically, there are inadequate stars to determine the variation of the foreground extinction within the small distance range, in particular for TMC at $\sim150$pc.

In addition, only the stars with $\BpRp{}>$0.15mag, i.e. $>5\sigma$, are selected to ensure the `jump' is significant in the given sightline. Because the `jump' is likely to appear at the edge of the cloud, we include the stars at an extended distance range of each molecular cloud (as shown in Table \ref{Cloudboundary}) both at the close and the far side by 100pc to guarantee that the jump at the edge of the cloud can be detected.

An MCMC analysis is performed to find the best set of parameters in Equation \ref{CE_all} under the priors of uniform distribution for all parameters, which maximize the likelihood defined as :
\begin{equation}\label{Likelihood}
L(\mathbf{x} \mid \theta) = \prod^{n}_{i=1} \frac{1}{2\pi\sqrt{\sigma}}exp(-\frac{(E_{i}-E(d_{i}\mid \mathbf{\theta} ))^{2}}{2\sigma_{E_{i}}^{2}})
\end{equation}
where $\theta$ is the parameter to be determined, i.e. $E_{0}$, $\delta E$, $ d^{MC}$ and $\sigma$; $E_{i}$ and $E(d_{i} \mid \theta)$ are the color excess calculated from the blue-edge method and the equation parameter to be fitted; $\sigma_{E_{i}}$ is the error of the color excess; $x$ is the data used to fit the equation, including $E_{i}$, $d_{i}$ and $\sigma_{E_{i}}$; and $n$ is the total number of stars in each pixel. We now consider the uncertainty of the distance and combine the distance uncertainty together with the uncertainty of the derived color excess as \citet{Chen19} (c.f. their Equation 2). With the assistance of the distance from \citet{Bjdis21}, the $\sigma_{E_{i}}$ in the likelihood function is given by:
\begin{equation}\label{CEerror}
\sigma_{E_{i}}^{2} = \sigma_{E^{obs}_{i}}^{2}+\sigma_{E^{BlueEdge}_{i}}^{2}+(E_{i}\frac{\sigma_{d_{i}}}{d_{i}})^{2}
\end{equation}
where $\sigma_{E^{obs}_{i}}$ is the uncertainty of the observed color index, $\sigma_{E^{BlueEdge}_{i}}$ is the uncertainty imported by the blue-edge method as discussed in Section \ref{SecBE}, $(E_{i}\frac{\sigma_{d_{i}}}{d_{i}})$ results from the distance uncertainty, which is only an approximation under the assumption that the dust opacity is constant along the line of sight. The distance uncertainties are simply adopted as $\sigma_{d_{i}} = \frac{d_{hi}-d_{lo}}{2}$ where $d_{hi}$ and $d_{lo}$ are the upper and lower bounds in the \citet{Bjdis21} distance catalog.

Stars that lie within the ranges of both the angular radius and the extended distance of the specific MC are adopted to fit Equation \ref{CE_all}, \ref{CE_mc} and \ref{CE_fg}. The MCMC procedure \citep{MCMC_F13} is performed to fit the parameters in the model. The `burn-in' chain has 50 walkers and 500 steps to stabilize the chain for final Monte Carlo simulation. Then 3000 steps with 50 walkers are run to estimate the final parameters and their errors. The median values (50th percentile) of the final chain are taken as the best estimation, and the uncertainties equal to the 16th and 84th percentile. Although the input distance range is extended, the result is required to satisfy the following condition so that the `jump' is still in the distance range of the molecular cloud:
\begin{equation}\label{dis_condition}
d^{\rm MC}_{\rm lower}<d^{\rm MC}-\sigma<d^{\rm MC}+\sigma<d^{\rm MC}_{\rm upper}
\end{equation}
where $d^{\rm MC}_{\rm lower}$ and $d^{\rm MC}_{\rm upper}$ mark the lower and upper limit of the MC's distance, $d^{\rm MC}$ and $\sigma$ mark the distance and the half thickness with error of the specific cloud region.

\begin{figure}
\centering
\centerline{\includegraphics[scale=0.6]{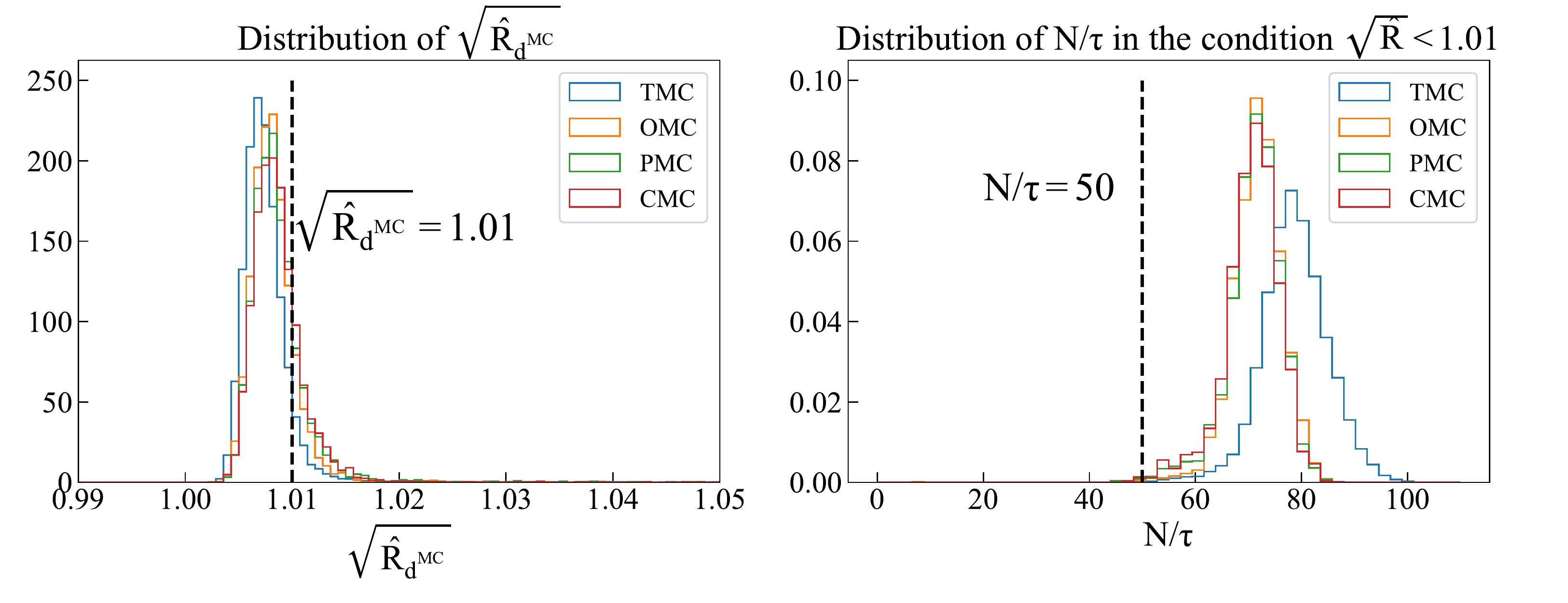}}
\caption{The left panel shows the $\sqrt{\hat{R}}$ of the parameter $d^{\rm {MC}}$ of the four MCs in the MCMC fitting to check the convergence. The right panel shows the distribution of effective samples in four MCs for the converged fittings with the criteria that the $\sqrt{\hat{R}}$ for all the parameter are smaller than 1.01.
\label{convergence}}
\end{figure}

We adopt the Gelman$-$Rubin statistic to determine whether the fit reaches to convergence. The MCMC fitting is regarded to be converged if the square root of the R hat $(\sqrt{\hat{R}})$ for all the parameters are smaller than 1.01 \citep{R_hat_19}. We first run the chain 2 times for each sightline and the parameter set with the smaller $\sqrt{\hat{R}}$ for the $d^{\rm {MC}}$ is adopted. If the value of $\sqrt{\hat{R}}$ is smaller than 1.01 for every parameter, this set of parameters is believed to converge and taken as the final results for the fitting in the given sightline, which leads to that about 55$\%$ of the fittings are converged. Moreover, the integrated autocorrelation time $\tau$ is used to check the effective number of independent samples for the converged fittings (typically the chains longer than about 50$\tau$ are sufficient \citep{MCMC_F13}). Figure \ref{convergence} shows the convergence results of the distance ($d^{\rm {MC}}$). The right panel in Figure \ref{convergence} indicates that almost all (about 99$\%$) of the converged fittings have sufficient independent samples, which strengthens the validity of results selected by R hat.

\begin{figure}
\centering
\centerline{\includegraphics[scale=0.5]{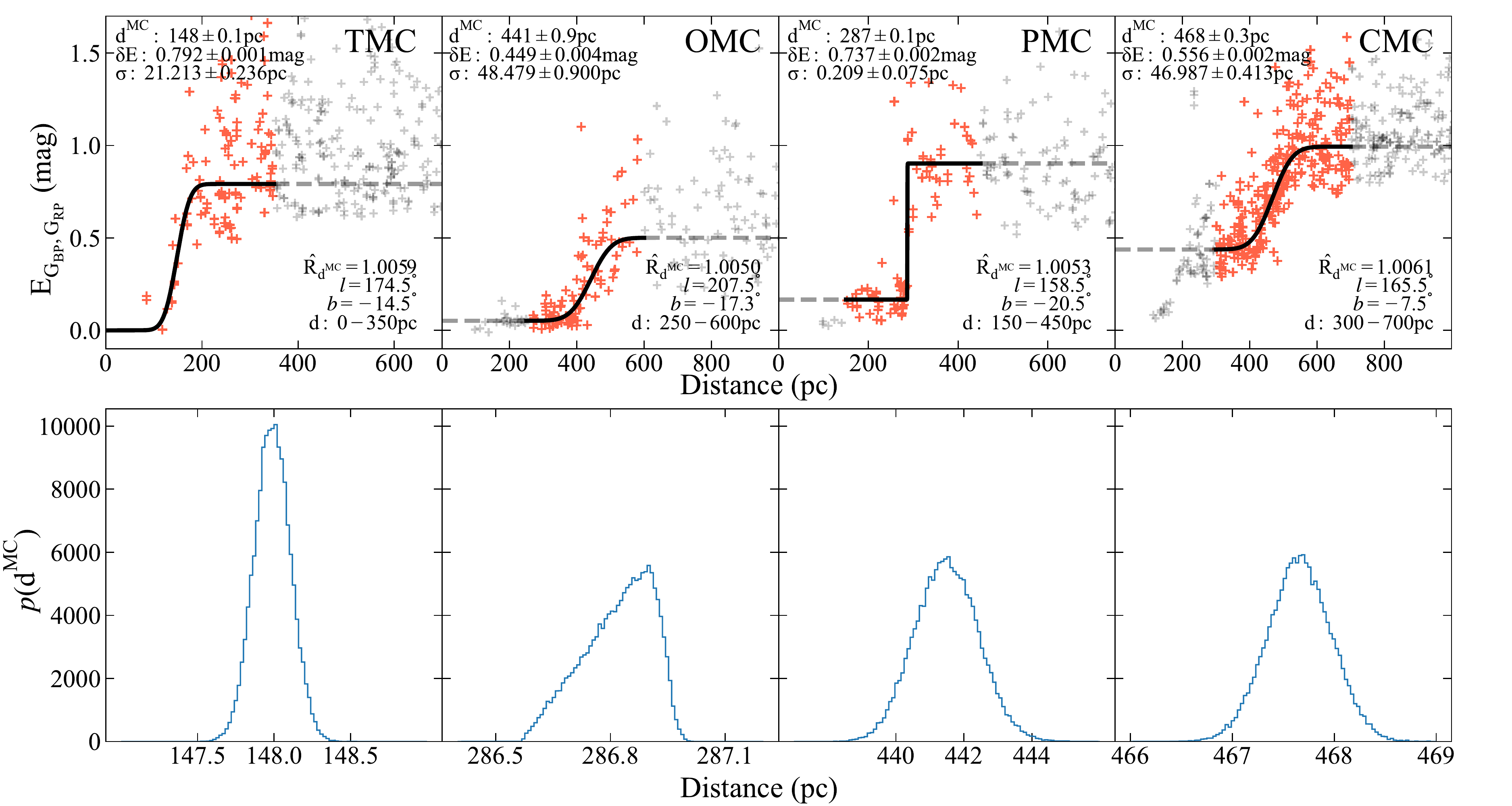}}
\caption{Four panels above show the fitting results (black solid line) of the extinction distance model (Equation \ref{CE_all}, \ref{CE_mc} and \ref{CE_fg}) for $\BpRp{}$ in four sightlines. The red cross marks the star that lies both within $1\degr$ of the sightline and within the distance range and is used to fit the model. The grey cross marks the star that lies in the sightline but is out of the distance range and thus not used in fitting. The fitted parameters of Equation \ref{CE_mc}  are  in the upper left, and the range is shown in the lower right. The four panels below show the distribution of the parameter $d^{MC}$ in posterior samples.
\label{mcmcsample}}
\end{figure}

Figure \ref{mcmcsample} displays four examples, each for one MC, of the fitting result, i.e. the $\hat{R}$ for the parameter distance ($d^{\rm {MC}}$) and the distribution of the posterior samples of $d^{\rm {MC}}$. It can be seen that the model well follows the trend of the observational points in the expected distance range and the fitting is well converged.

\begin{figure}
\centering
\centerline{\includegraphics[scale=0.5]{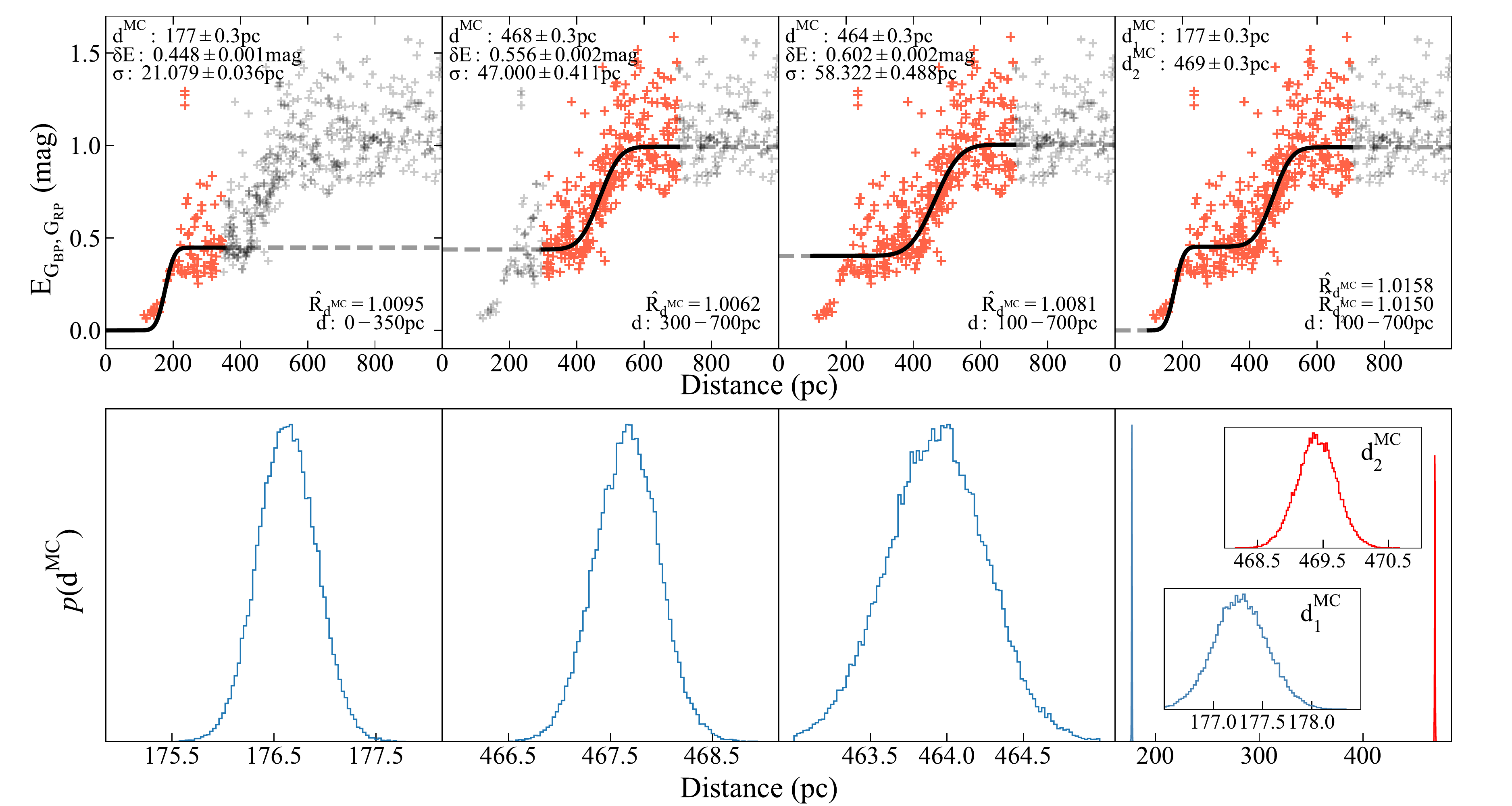}}
\caption{Comparison of the one-cloud model with preselected distance range and the two-cloud model with no distance limitation, the inset illustrates the distribution of two distance parameters in two-cloud model individually.
\label{muticloudsample}}
\end{figure}

In some sightline, one-cloud model is not perfectly optimized to fit the MCs, there are two or more clouds in some sightlines. We test the two-cloud system with four models as an example, the case sightline is the fourth example in Figure \ref{mcmcsample} which transverses both the TMC and CMC clouds. In Figure \ref{muticloudsample}, four models are run: (1) The distance range is limited from 100pc to 350pc and one-cloud model is used, then the distance is derived and converged, which detects the TMC cloud; (2) the distance is limited from 300pc to 700pc and one-cloud model is used, then we obtain the distance of convergence for the CMC cloud; (3) the distance is limited from 100 to 700pc and one-cloud model is used, then the result will be similar to the result from 300pc to 700pc, which detects the distance of CMC with the higher extinction and more star tracers; and (4) the distance is again limited from 100pc to 700pc but two-cloud model is used, then two distances, i.e. the distance to the TMC and CMC cloud, are detected, which completely coincide with the one-cloud model with a preselected distance range. This proves that the one-cloud model with a preselected distance range yields the same results as a two-cloud model with no preselection of distance. Also, from the third model which is a two-cloud system but fits with one-cloud model, we find that the one-cloud model will detect the highest extinction cloud or the cloud containing more star tracers if directly used to fit a two-cloud system. The clouds are complexes and may contain multiple cloudlets. Such complexes are presented within a large space range, and the sub-structures mostly disperse in distance. In our analysis, the bin-size in space is only $0.2^\circ$ by side, in which we consider no complexes. Indeed, a small distance separation (say less than 5pc) will be difficult to distinguish with the present accuracy.

\section{RESULTS AND DISCUSSIONS}\label{SecRes}

\subsection{The extinction structure of individual molecular cloud} \label{3DMC}

With the distance and the sliced extinction determined, the 2D extinction map can be decomposed into the individual map for each MC. In Figure \ref{MapTau}, \ref{MapPer}, \ref{MapOri} and \ref{MapCali} for the four clouds, the left and right panels are for $\BpRp{}$ and  $\JK{}$ respectively, while the middle panels show the distance structure with the background contour map of $\BpRp{}$. The blanks denote the position with no reliable result.

\subsubsection{Taurus}\label{tmc_result}

\begin{figure}
\centering
\centerline{\includegraphics[scale=0.6]{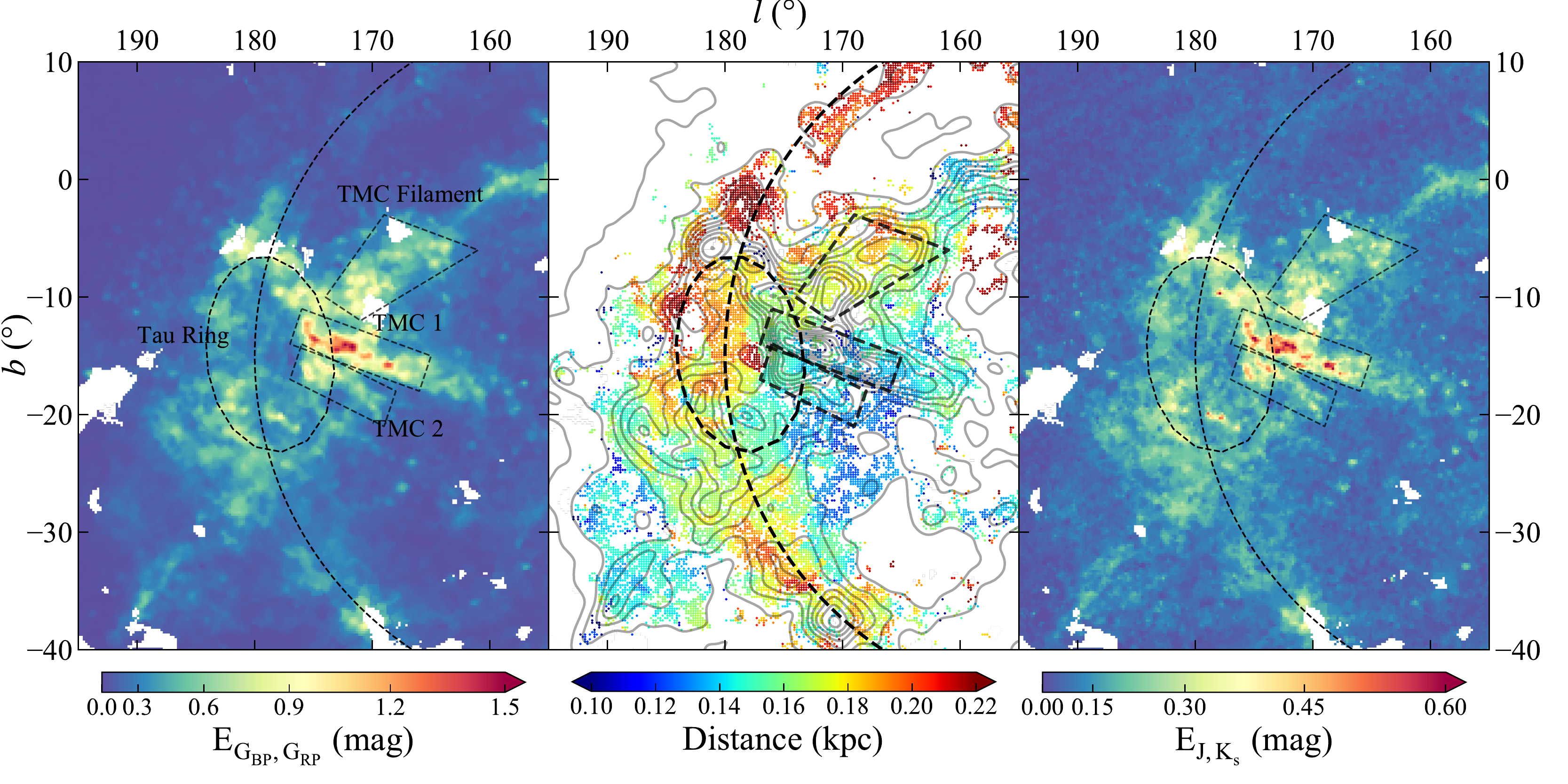}}
\caption{The extinction map of TMC in $\BpRp{}$ (left) and $\JK{}$ (right) integrated from 0pc to 250pc. The main sub-structures (TMC Filament, TMC1, TMC2, Tau Ring and the bow-like structure) are marked by the black dashed line.  The color in the middle panel denotes the distance in each pixel, and the contour map denotes $\BpRp{}$.
\label{MapTau}}
\end{figure}

The extinction map of TMC is integrated over distance from 0 to 250pc and displayed in Figure \ref{MapTau}. TMC has four prominent substructures, i.e. TMC1 and TMC2 \citep{TauCaliPer2010III} being the mostly studied regions, the Tau Ring \citep{localMC21} and the TMC filament. These four substructures are clearly visible in the extinction map (Figure \ref{MapTau}), where the boundaries of TMC1 and TMC2 are taken from \citet{TauCaliPer2010III} and \citet{OCTP4cloud}, and the Tau Ring and the filament are plotted according to \citet{TPshell21}. Consistent with previous studies, TMC1 presents the most serious extinction that its densest position has an  $\BpRp{} > 1.5$mag or $\JK{} > 0.6$mag. Similarly, TMC2 also has high extinction, though smaller than TMC1. From the extinction map, it can be seen that some positions within TMC2 also have $\JK{} > 0.6$mag. Differently, the Tau ring and filament are not so dense, but have a moderate extinction  mostly with $\BpRp{} < 1.0$mag. Accordingly, their structure looks much more diffuse than TMC1 and TMC2.

In terms of distance, TMC1 and TMC2 are close to each other in that TMC1 extends from 129pc to 157pc and TMC2 extends from 132pc to 156pc. This result coincides with that of \citet{Tau_zitai} who found the average distance of TMC to be $145_{-16}^{+12}$pc, and also agrees with the result of \citet{localMC21} that found TMC extends from 131pc to 168pc. Meanwhile, the TMC filament and the Tau Ring are at comparable distance around 174pc, apparently further than TMC1 and TMC2. But these structures are connected. Specifically, both TMC1 and TMC2 extend to further distance with the ascending of longitude and finally connect to the Tau Ring at the edges. Independently, the Tau Ring can be depicted by an ellipse as suggested by \citet{TPshell21},  whose semi-major and semi-minor axis are 39pc and 26pc respectively (c.f. Figure 5 in \citealt{TPshell21}). The center of the ellipse locates at $(l,b,d) = (179\degr.5,\ -14\degr.2,\ 179
{\rm pc})$. The distance of the Tau Ring extends from $\sim$150pc to 220pc where the closer side lies at the higher latitude and the further side at the lower latitude. The TMC filament extends to the midplane area within a distance near 174pc. Part of this filament overlaps with the CMC in sightline. Such overlapping can be seen in the fourth panel of Figure \ref{mcmcsample} where two `jumps' are visible, the first `jump' is induced by the TMC filament at about 174pc and the second `jump' is induced by CMC at around 480pc. Fortunately, the distance can separate them unambiguously in this work.

In addition to the four sub-structures that belong to TMC, there is a large-scale bow-like structure which appears in the extinction slice graph (Figure \ref{slice}) from 175pc to 250pc which the Tau ring belongs to. The discussion of the shell and the bow will be presented in Section \ref{LB_TP}.

\subsubsection{Perseus}\label{pmc_result}

\begin{figure}
\centering
\centerline{\includegraphics[scale=0.6]{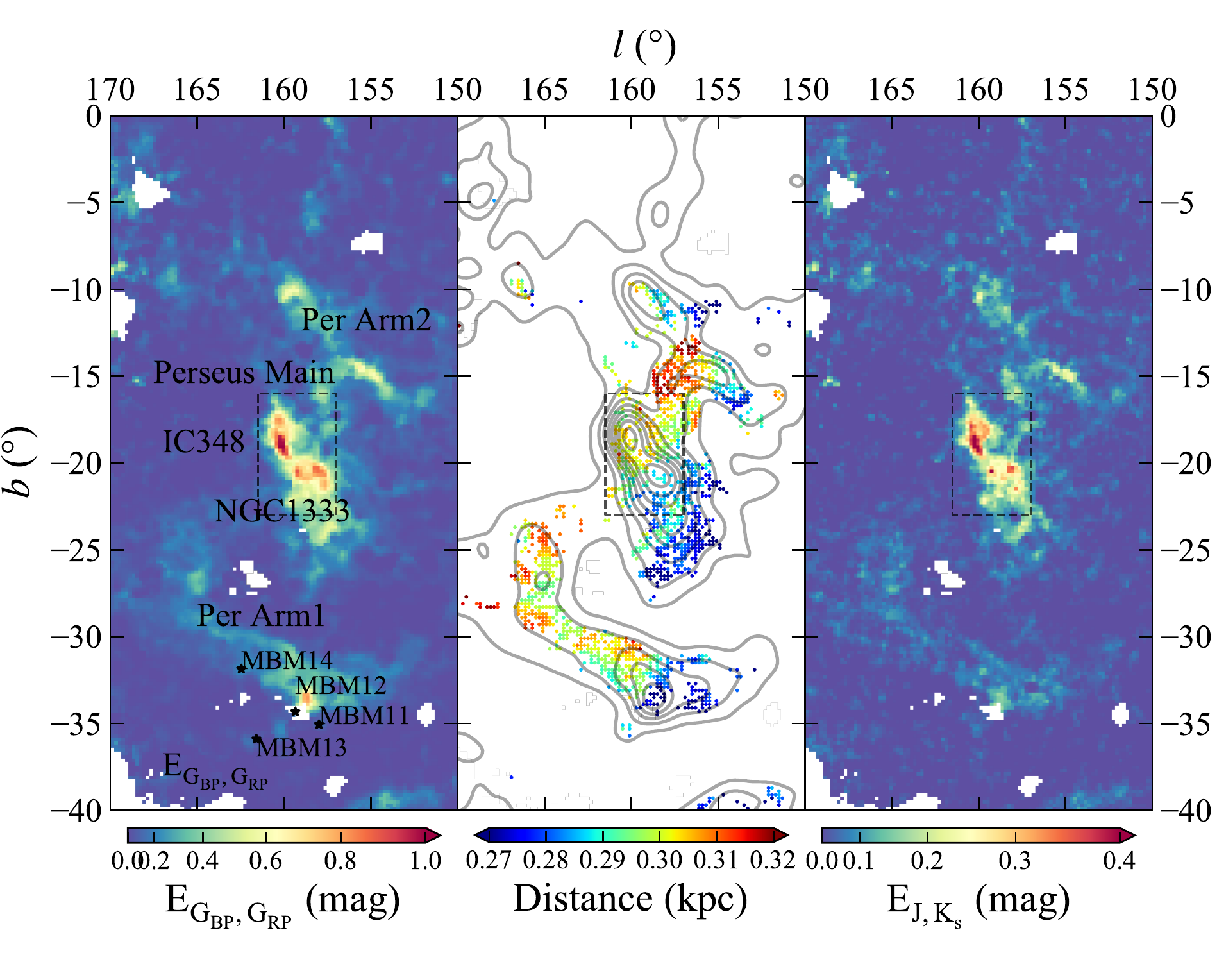}}
\caption{The extinction map of PMC in $\BpRp{}$ (left) and $\JK{}$ (right) integrated from 250pc to 350pc. The symbols follow the convention in Figure \ref{MapTau}.
\label{MapPer}}
\end{figure}

The extinction map of PMC is integrated over the distance from 250pc to 350pc, and displayed in Figure \ref{MapPer}. Consistent with previous studies, the Perseus Main structure is obvious. There are two other substructures that appear in a few 3D extinction maps, e.g. \citet{L20} and filamentary structure in \citet{OCTP4cloud}. Here, they are clearly recognizable in the middle panel of Figure \ref{MapPer} in that their distances increase with the longitude, the same as the Perseus Main. In addition, the distribution of extinction is continuous to the Perseus Main. Thus we think they are part of PMC and name them Per Arm1 and Per Arm2 respectively.

In comparison with TMC, PMC is not so dense. The most serious extinction occurs in the Main part, where the largest color excess $\BpRp{}$ is about $1.0$mag. The Arm1 and Arm2 have an extinction of about $\BpRp{} \sim 0.4$mag, i.e. $A_{\rm V} \sim 1.0$mag, consistent with the feature of a translucent molecular cloud. Indeed, the extension of the cloud in the radial direction is only about 30pc, which can partly account for the relatively low extinction.

There are two famous clusters located within the Perseus Main, i.e. IC 348 ($(l,b) = (160\degr.50,\ -18\degr.27)$) and NGC 1333 ($(l,b) = (158\degr.34,\ -20\degr.64)$) \citep{Per_zucker18}. Figure \ref{MapPer} shows an obvious distance gradient of the PMC Main that gradually becomes further from bottom to top within a range of $\sim$285pc to $\sim$306pc, inferring a thickness of 30pc for this cloud. The distance to IC 348 and NGC 1333 is 304pc and 287pc respectively. This result agrees with that of \citet{Per_zucker18} who suggested $295\pm4$pc and $299\pm3$pc for IC 348 and NGC 1333 respectively. However, \citet{Per_Go_YSO18} recommended a distance of $321\pm27$pc and $294\pm28$pc for IC 348 and NGC 1333 respectively from the Gaia data. But their results bear a large uncertainty possibly due to the dispersion of parallaxes of YSOs, though the results are still consistent within the uncertainties.

A few high latitude molecular clouds including MBM\,11-14 \citep{MBMlist85} are in the area of the Perseus Arm1 in the left panel of Figure \ref{MapPer}. \citet{Sun21_MCdistance} derived the distance to these four clouds to be 147pc for MBM\,11, 278pc for MBM\,12, 409pc for MBM\,13, and 295pc for MBM\,14, which implies that only MBM\,12 and MBM\,14  are associated with the Perseus Cloud, while MBM\,11 is in front of the cloud and MBM\,13 is behind the cloud. The distribution of $\BpRp{}$ looks consistent with this result in that the integrated extinction from 250-350pc is small for MBM\,11 and MBM\,13.

\citet{OCTP4cloud} named some filamentary structures such as the California, Taurus and Perseus Filament that are visible in Figure 12 of \citet{OCTP4cloud}. Compared with our work, Per Arm2 and the California filament similar structure in that they share the approximate coordinates and distance. Specifically, Per Arm2 ranges from about 290pc to 300pc while the California filament is from about 250pc to 350pc. Apart from this, the Taurus and Perseus Filament may not be the same structure as Perseus Arm1, since the angular sizes of the Taurus and Perseus filament are only about several degrees while Perseus Arm1 has an angular size of more than $10\degr$. Besides, the field in Figure 12 of \citet{OCTP4cloud} is within about $154\degr<l<164\degr$ and $-25\degr<b<-15\degr$, while the majority of Perseus Arm1 is higher than $-25\degr$ in the Galactic latitude.

\subsubsection{Orion}\label{omc_result}

\begin{figure}
\centering
\centerline{\includegraphics[scale=0.6]{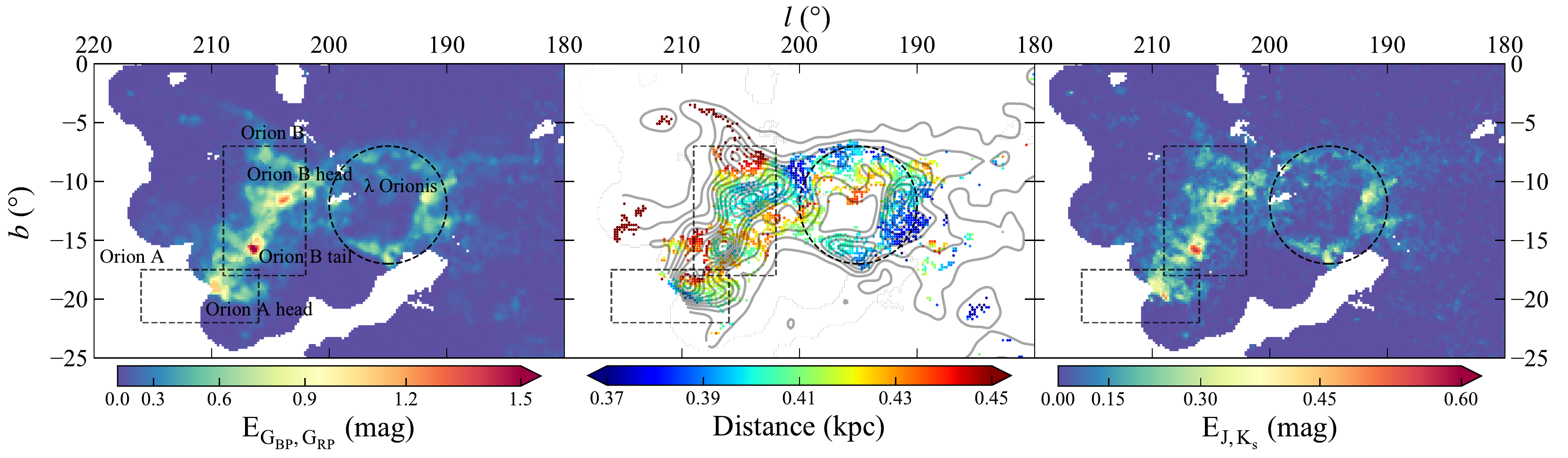}}
\caption{The extinction map of OMC in $\BpRp{}$ (left) and $\JK{}$ (right) integrated from 350pc to 500pc. The symbols follow the convention in Figure \ref{MapTau}.
\label{MapOri}}
\end{figure}

The extinction map of OMC is integrated over distance from 350pc to 500pc, and displayed in Figure \ref{MapOri}. OMC is usually divided into three parts according to the position and the morphology in the extinction map, i.e. Orion A, Orion B and $\lambda$ Orionis. Due to the data limits, only the head part is investigated for Orion A. The boundaries of each part in Figure \ref{MapOri} are taken from \citet{Orion2011IV} and \citet{OCTP4cloud}. Orion B exhibits the most serious extinction, where the densest extinction of the head and the tail can reach up to $\BpRp{} > 1.5$mag. The observable part of Orion A is much less obscured, and $\lambda$ Orionis is even more diffuse with a color excess of about $\BpRp{} \sim 0.3-0.5$mag.

The closest part of OMC is $\lambda$ Orionis, for which a clear ring-like structure is evident in the extinction map. Figure \ref{MapOri} shows $\lambda$ Orionis stretches from $\sim$380pc to $\sim$400pc, which agrees with the range from 375pc to 397pc by \citet{localMC21}. The ring can be divided into two halves, the lower half extends from 380pc to 400pc, while the upper half is connected with the Orion B head, which is at a distance around 410pc. For Orion B, the close part is the head at 410pc with $b=-10^{\circ}$ and extends towards the tail, with a distance slowly increasing to 420pc. For the head of Orion A, the distance is about 420pc, which is larger than $393\pm 25$pc in \citet{OrionA_YSO_18} by using YSOs as tracers. Overall, Orion A and B are at comparable distance and slightly further than $\lambda$ Orionis, and OMC is not so extended in the radial direction as TMC.

\subsubsection{California}\label{cmc_result}
\begin{figure}
\centering
\centerline{\includegraphics[scale=0.6]{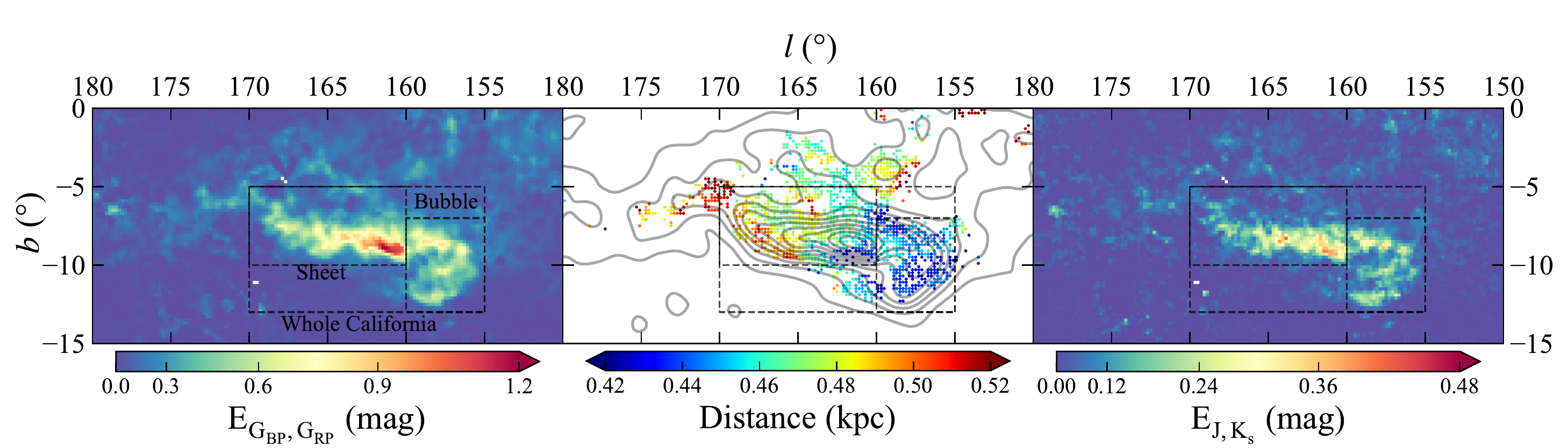}}
\caption{The extinction map of CMC in $\BpRp{}$ (left) and $\JK{}$ (right) integrated from 400pc to 600pc. The symbols follow the convention in Figure \ref{MapTau}.
\label{MapCali}}
\end{figure}

The extinction map of CMC is integrated over distance from 400-600pc, and displayed in Figure \ref{MapCali}. The densest position has $\BpRp{}>1.2$mag, slightly higher than PMC, though it still resembles a translucent other than dense molecular cloud.

As shown in Figure \ref{MapCali}, the structure of CMC is comparatively simple. It is a sheet structure extending from $\sim$ 448pc to $\sim$504pc with $(l,b) = ([160\degr, 170\degr],\ [-10\degr, -5\degr])$. A bubble structure (the name is from \citealt{CaliOriA21_Kh}) exists with $(l,b) = ([155\degr, 160\degr],\ [-13\degr, -7\degr])$ in the extinction map at a distance from 440pc to 448pc. \citet{CaliOriA21_Kh} presents a 3D view of CMC in their Figure 2, in which the dense region of the bubble is on the small longitude and is apparently seen around 455pc, and the filament on the large longitude is around 495pc to 515pc, which is in agreement with this work.

\subsection{The Shell-Like Structure}\label{Shell}

\subsubsection{The Per-Tau Shell and a Bow Like Structure}\label{LB_TP}

\citet{TPshell21} revealed the Per-Tau Shell that is an extended near-spherical shell embedding in PMC and TMC. They discuss a scenario that the ISM swept up by supernova and stellar feedback events forms the expanding shell containing both TMC and PMC. They suggest that Per-Tau Shell looks like a circle with the center at $(l,b,d) = (161\degr.1,\ -22\degr.7,\ 218{\rm pc})$ and a radius of 78pc.

The bow-like structure is most obvious in the 200-225pc slice in Figure \ref{slice}, while visible from $d=175$pc to $d=250$pc. Thus the extinction is integrated from 175pc to 250pc to increase the visibility and shown in the left panel of Figure \ref{RingBow}, where the identified shell is marked by the red dashed line and the Per-Tau shell is indicated by the black dashed line. It can be seen that the lower part of the shell coincides with the Per-Tau Shell in both the coordinates and the distance. At the top, the Per-Tau Shell is consistent with the TMC filament visible in the 175-200pc slice, while the bow structure identified in this work stretches to another filament that is at higher latitude than the TMC filament and visible in the 200-225pc slice. Overall, the bow-like structure around 180-220pc is further than the main substructures of TMC, i.e. TMC1 and TMC2 around 140-160pc, therefore it can be regarded as an independent structure.

\subsubsection{The Low-extinction Rings} \label{R1R2}

\begin{figure}
\centering
\centerline{\includegraphics[scale=0.65]{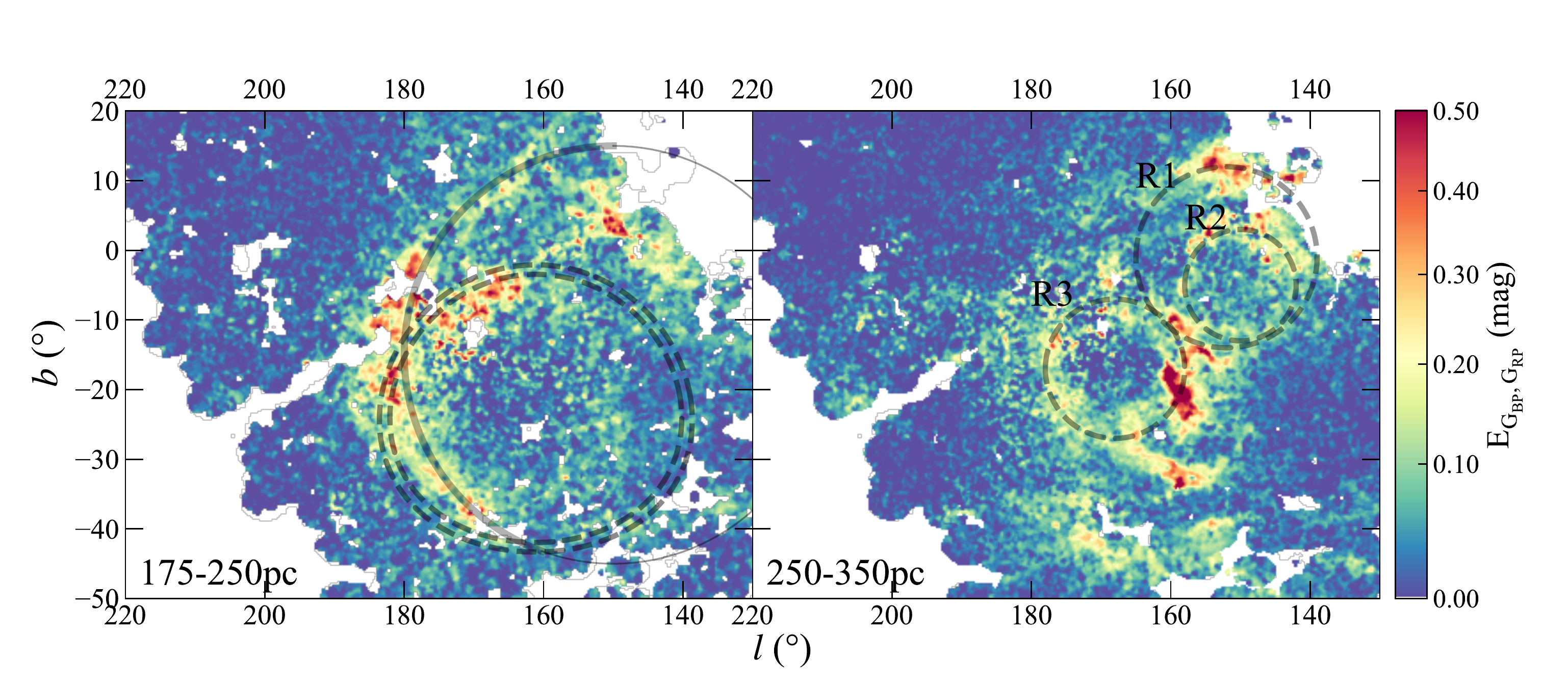}}
\caption{Left: the bow-like structure (gray solid arc) in $\BpRp{}$ integrated from 175pc to 250pc and the Per-Tau shell (gray dashed ring). Right: three possible ring-like structures (gray dashed line) in $\BpRp{}$ integrated from 250pc to 350pc.}
\label{RingBow}
\end{figure}

Three other ring-like structures are revealed in the extinction map and shown by the red dashed circles in the right panel of Figure \ref{RingBow}. They are further than the known Per-Tau shell and the extinction is integrated from 250pc to 350pc. Indeed, they are at the same distance range as PMC, so the two high-extinction structures in Figure \ref{RingBow} belong to PMC other than the ring. Excluding the PMC structures, the center of the projection is located approximately at $(l,b) = (152\degr,\ -1\degr)$ with a radius of 13\degr for R1, $(l,b) = (150\degr,\ -5\degr)$ with a radius of 8\degr\ for R2, and $(l,b) = (168\degr,\ -17\degr)$ with a radius of 10\degr\ for R3.  R1 and R2 are visible in the 250pc to 300pc slice, and R3 is visible in the 300pc to 350pc slices in Figure \ref{slice}. For a given distance of 300pc, the linear radius of the three rings is about 68pc, 52pc, and 42pc for R1, R2 and R3 respectively, which is compatible with the size of an old supernova remnant. Moreover, the color excess in the main part of the rings is only $\sim 0.1$mag (at the $3\sigma$ level) to $0.2$mag in $\BpRp{}$ i.e. $A_{V}\sim $ $0.2$mag to $0.5$mag, smaller than the above bow-like structure, while consistent with an old supernova remnant. However, they are not in the Green's list of supernova remnants \citep{SNRcata_Green19}. The further identification of the rings is interesting but beyond the scope of this work.

The parameters of all the substructures identified in the four clouds are summarized in Table \ref{MCsTable}.

\section{Summary} \label{Summary}

The extinction structure is studied by high-precision color excesses of the stars in the sky area of the Taurus, Orion, Perseus and California molecular cloud. The intrinsic color indexes are derived by the blue-edge method from the atmospheric parameters obtained by the LAMOST spectroscopic survey, and the observed ones are calculated from the Gaia and 2MASS photometry in the $G_{\rm BP}$, $G_{\rm RP}$, $J$ and $K_{\rm s}$ bands. The resultant error is about $\sim$0.03mag and $\sim$0.07mag for $\BpRp{}$ and $\JK{}$ respectively.

In combination with the distance measured by Gaia, the distance-sliced extinction map at a step of 25pc is built up by assuming that the extinction is monotonically increasing with distance. It well separates the clouds by the distance and delimits the range of each cloud in 3D space. In addition, the distance to each cloud segment is more accurately determined from the extinction-jump model, i.e. the extinction increases sharply at the distance of the cloud segment. The extinction map is then yielded by integrating the extinction over the distance range of the cloud, which includes some low-extinction regions. The extinction structure confirms the previously identified dense sub-structure like TMC1, TMC2, Tau Ring, Orion A, Orion B, $\lambda$ Orionis, and Perseus Main. It also finds additional structures. Two arms in the Perseus cloud (Perseus Arm2 is similar to the California filament in \citet{OCTP4cloud}) are identified for their geometrical connection with the Per Main and evident extinction. A bow-like structure is presented at a distance around 200pc, which overlaps partly with the Per-Tau shell but deviates from it at about $b=-$10$\degr$. Three new rings are visible at the level of $\BpRp{} \sim 0.1-0.2$mag. The stellar color excesses and the extinction maps will be used in future work to study the extinction law in the star-forming regions and its dependence on the environment.

\acknowledgments{We are grateful to Drs. Jian Gao, Haibo Yuan, Jun Li, Shu Wang, Cunying Xiao and Mr. Tianding Wang for their friendly help and discussion. We thank the anonymous referee for very useful suggestions to improve the work. This work is supported by the NSFC projects 12133002 and 12203016, National Key R\&D Program of China No.2019YFA0405503, CMS-CSST-2021-A09, Natural Science Foundation of Hebei Province (No.A2022205018) and Science Foundation of Hebei Normal University (No.L2022B33). This work has made use of the data from LAMOST, Gaia and 2MASS.}

\software{scikit-learn \citep{scikit-learn}, emcee \citep{MCMC_F13}, dustmaps \citep{dustmaps}.}

\clearpage
\bibliographystyle{aasjournal}
\bibliography{main}

\begin{thebibliography}{}
\expandafter\ifx\csname natexlab\endcsname\relax\def\natexlab#1{#1}\fi
\providecommand{\url}[1]{\href{#1}{#1}}
\providecommand{\dodoi}[1]{doi:~\href{http://doi.org/#1}{\nolinkurl{#1}}}
\providecommand{\doeprint}[1]{\href{http://ascl.net/#1}{\nolinkurl{http://ascl.net/#1}}}
\providecommand{\doarXiv}[1]{\href{https://arxiv.org/abs/#1}{\nolinkurl{https://arxiv.org/abs/#1}}}

\bibitem[{{Bailer-Jones} {et~al.}(2021){Bailer-Jones}, {Rybizki}, {Fouesneau},
  {Demleitner}, \& {Andrae}}]{Bjdis21}
{Bailer-Jones}, C.~A.~L., {Rybizki}, J., {Fouesneau}, M., {Demleitner}, M., \&
  {Andrae}, R. 2021, \aj, 161, 147, \dodoi{10.3847/1538-3881/abd806}

\bibitem[{{Bialy} {et~al.}(2021){Bialy}, {Zucker}, {Goodman}, {Foley}, {Alves},
  {Semenov}, {Benjamin}, {Leike}, \& {En{\ss}lin}}]{TPshell21}
{Bialy}, S., {Zucker}, C., {Goodman}, A., {et~al.} 2021, \apjl, 919, L5,
  \dodoi{10.3847/2041-8213/ac1f95}

\bibitem[{{Chen} {et~al.}(2017){Chen}, {Liu}, {Ren}, {Yuan}, {Huang}, {Yu},
  {Xiang}, {Wang}, {Tian}, \& {Zhang}}]{Chen17_s147}
{Chen}, B.~Q., {Liu}, X.~W., {Ren}, J.~J., {et~al.} 2017, \mnras, 472, 3924,
  \dodoi{10.1093/mnras/stx2287}

\bibitem[{{Chen} {et~al.}(2019){Chen}, {Huang}, {Yuan}, {Wang}, {Fan}, {Xiang},
  {Zhang}, {Tian}, \& {Liu}}]{Chen19}
{Chen}, B.~Q., {Huang}, Y., {Yuan}, H.~B., {et~al.} 2019, \mnras, 483, 4277,
  \dodoi{10.1093/mnras/sty3341}

\bibitem[{{Danielski} {et~al.}(2018){Danielski}, {Babusiaux}, {Ruiz-Dern},
  {Sartoretti}, \& {Arenou}}]{Dan18_G_extinction}
{Danielski}, C., {Babusiaux}, C., {Ruiz-Dern}, L., {Sartoretti}, P., \&
  {Arenou}, F. 2018, \aap, 614, A19, \dodoi{10.1051/0004-6361/201732327}

\bibitem[{{Dharmawardena} {et~al.}(2022){Dharmawardena}, {Bailer-Jones},
  {Fouesneau}, \& {Foreman-Mackey}}]{OCTP4cloud}
{Dharmawardena}, T.~E., {Bailer-Jones}, C.~A.~L., {Fouesneau}, M., \&
  {Foreman-Mackey}, D. 2022, \aap, 658, A166,
  \dodoi{10.1051/0004-6361/202141298}

\bibitem[{{Ducati} {et~al.}(2001){Ducati}, {Bevilacqua}, {Rembold}, \&
  {Ribeiro}}]{Jorge_2001_IRcolor}
{Ducati}, J.~R., {Bevilacqua}, C.~M., {Rembold}, S.~B., \& {Ribeiro}, D. 2001,
  \apj, 558, 309, \dodoi{10.1086/322439}

\bibitem[{{Foreman-Mackey} {et~al.}(2013){Foreman-Mackey}, {Hogg}, {Lang}, \&
  {Goodman}}]{MCMC_F13}
{Foreman-Mackey}, D., {Hogg}, D.~W., {Lang}, D., \& {Goodman}, J. 2013, \pasp,
  125, 306, \dodoi{10.1086/670067}

\bibitem[{{Gaia Collaboration} {et~al.}(2016){Gaia Collaboration}, {Prusti},
  {de Bruijne}, {Brown}, {Vallenari}, {Babusiaux}, {Bailer-Jones}, {Bastian},
  {Biermann}, {Evans}, {Eyer}, {Jansen}, {Jordi}, {Klioner}, {Lammers},
  {Lindegren}, {Luri}, {Mignard}, {Milligan}, {Panem}, {Poinsignon},
  {Pourbaix}, {Randich}, {Sarri}, {Sartoretti}, {Siddiqui}, {Soubiran},
  {Valette}, {van Leeuwen}, {Walton}, {Aerts}, {Arenou}, {Cropper}, {Drimmel},
  {H{\o}g}, {Katz}, {Lattanzi}, {O'Mullane}, {Grebel}, {Holland}, {Huc},
  {Passot}, {Bramante}, {Cacciari}, {Casta{\~n}eda}, {Chaoul}, {Cheek}, {De
  Angeli}, {Fabricius}, {Guerra}, {Hern{\'a}ndez}, {Jean-Antoine-Piccolo},
  {Masana}, {Messineo}, {Mowlavi}, {Nienartowicz}, {Ord{\'o}{\~n}ez-Blanco},
  {Panuzzo}, {Portell}, {Richards}, {Riello}, {Seabroke}, {Tanga},
  {Th{\'e}venin}, {Torra}, {Els}, {Gracia-Abril}, {Comoretto},
  {Garcia-Reinaldos}, {Lock}, {Mercier}, {Altmann}, {Andrae}, {Astraatmadja},
  {Bellas-Velidis}, {Benson}, {Berthier}, {Blomme}, {Busso}, {Carry},
  {Cellino}, {Clementini}, {Cowell}, {Creevey}, {Cuypers}, {Davidson}, {De
  Ridder}, {de Torres}, {Delchambre}, {Dell'Oro}, {Ducourant}, {Fr{\'e}mat},
  {Garc{\'\i}a-Torres}, {Gosset}, {Halbwachs}, {Hambly}, {Harrison}, {Hauser},
  {Hestroffer}, {Hodgkin}, {Huckle}, {Hutton}, {Jasniewicz}, {Jordan},
  {Kontizas}, {Korn}, {Lanzafame}, {Manteiga}, {Moitinho}, {Muinonen},
  {Osinde}, {Pancino}, {Pauwels}, {Petit}, {Recio-Blanco}, {Robin}, {Sarro},
  {Siopis}, {Smith}, {Smith}, {Sozzetti}, {Thuillot}, {van Reeven}, {Viala},
  {Abbas}, {Abreu Aramburu}, {Accart}, {Aguado}, {Allan}, {Allasia},
  {Altavilla}, {{\'A}lvarez}, {Alves}, {Anderson}, {Andrei}, {Anglada Varela},
  {Antiche}, {Antoja}, {Ant{\'o}n}, {Arcay}, {Atzei}, {Ayache}, {Bach},
  {Baker}, {Balaguer-N{\'u}{\~n}ez}, {Barache}, {Barata}, {Barbier}, {Barblan},
  {Baroni}, {Barrado y Navascu{\'e}s}, {Barros}, {Barstow}, {Becciani},
  {Bellazzini}, {Bellei}, {Bello Garc{\'\i}a}, {Belokurov}, {Bendjoya},
  {Berihuete}, {Bianchi}, {Bienaym{\'e}}, {Billebaud}, {Blagorodnova},
  {Blanco-Cuaresma}, {Boch}, {Bombrun}, {Borrachero}, {Bouquillon}, {Bourda},
  {Bouy}, {Bragaglia}, {Breddels}, {Brouillet}, {Br{\"u}semeister},
  {Bucciarelli}, {Budnik}, {Burgess}, {Burgon}, {Burlacu}, {Busonero}, {Buzzi},
  {Caffau}, {Cambras}, {Campbell}, {Cancelliere}, {Cantat-Gaudin}, {Carlucci},
  {Carrasco}, {Castellani}, {Charlot}, {Charnas}, {Charvet}, {Chassat},
  {Chiavassa}, {Clotet}, {Cocozza}, {Collins}, {Collins}, {Costigan}, {Crifo},
  {Cross}, {Crosta}, {Crowley}, {Dafonte}, {Damerdji}, {Dapergolas}, {David},
  {David}, {De Cat}, {de Felice}, {de Laverny}, {De Luise}, {De March}, {de
  Martino}, {de Souza}, {Debosscher}, {del Pozo}, {Delbo}, {Delgado},
  {Delgado}, {di Marco}, {Di Matteo}, {Diakite}, {Distefano}, {Dolding}, {Dos
  Anjos}, {Drazinos}, {Dur{\'a}n}, {Dzigan}, {Ecale}, {Edvardsson}, {Enke},
  {Erdmann}, {Escolar}, {Espina}, {Evans}, {Eynard Bontemps}, {Fabre},
  {Fabrizio}, {Faigler}, {Falc{\~a}o}, {Farr{\`a}s Casas}, {Faye}, {Federici},
  {Fedorets}, {Fern{\'a}ndez-Hern{\'a}ndez}, {Fernique}, {Fienga}, {Figueras},
  {Filippi}, {Findeisen}, {Fonti}, {Fouesneau}, {Fraile}, {Fraser}, {Fuchs},
  {Furnell}, {Gai}, {Galleti}, {Galluccio}, {Garabato}, {Garc{\'\i}a-Sedano},
  {Gar{\'e}}, {Garofalo}, {Garralda}, {Gavras}, {Gerssen}, {Geyer}, {Gilmore},
  {Girona}, {Giuffrida}, {Gomes}, {Gonz{\'a}lez-Marcos},
  {Gonz{\'a}lez-N{\'u}{\~n}ez}, {Gonz{\'a}lez-Vidal}, {Granvik}, {Guerrier},
  {Guillout}, {Guiraud}, {G{\'u}rpide}, {Guti{\'e}rrez-S{\'a}nchez}, {Guy},
  {Haigron}, {Hatzidimitriou}, {Haywood}, {Heiter}, {Helmi}, {Hobbs},
  {Hofmann}, {Holl}, {Holland}, {Hunt}, {Hypki}, {Icardi}, {Irwin}, {Jevardat
  de Fombelle}, {Jofr{\'e}}, {Jonker}, {Jorissen}, {Julbe}, {Karampelas},
  {Kochoska}, {Kohley}, {Kolenberg}, {Kontizas}, {Koposov}, {Kordopatis},
  {Koubsky}, {Kowalczyk}, {Krone-Martins}, {Kudryashova}, {Kull}, {Bachchan},
  {Lacoste-Seris}, {Lanza}, {Lavigne}, {Le Poncin-Lafitte}, {Lebreton},
  {Lebzelter}, {Leccia}, {Leclerc}, {Lecoeur-Taibi}, {Lemaitre}, {Lenhardt},
  {Leroux}, {Liao}, {Licata}, {Lindstr{\o}m}, {Lister}, {Livanou}, {Lobel},
  {L{\"o}ffler}, {L{\'o}pez}, {Lopez-Lozano}, {Lorenz}, {Loureiro},
  {MacDonald}, {Magalh{\~a}es Fernandes}, {Managau}, {Mann}, {Mantelet},
  {Marchal}, {Marchant}, {Marconi}, {Marie}, {Marinoni}, {Marrese},
  {Marschalk{\'o}}, {Marshall}, {Mart{\'\i}n-Fleitas}, {Martino}, {Mary},
  {Matijevi{\v{c}}}, {Mazeh}, {McMillan}, {Messina}, {Mestre}, {Michalik},
  {Millar}, {Miranda}, {Molina}, {Molinaro}, {Molinaro}, {Moln{\'a}r},
  {Moniez}, {Montegriffo}, {Monteiro}, {Mor}, {Mora}, {Morbidelli}, {Morel},
  {Morgenthaler}, {Morley}, {Morris}, {Mulone}, {Muraveva}, {Musella},
  {Narbonne}, {Nelemans}, {Nicastro}, {Noval}, {Ord{\'e}novic},
  {Ordieres-Mer{\'e}}, {Osborne}, {Pagani}, {Pagano}, {Pailler}, {Palacin},
  {Palaversa}, {Parsons}, {Paulsen}, {Pecoraro}, {Pedrosa}, {Pentik{\"a}inen},
  {Pereira}, {Pichon}, {Piersimoni}, {Pineau}, {Plachy}, {Plum}, {Poujoulet},
  {Pr{\v{s}}a}, {Pulone}, {Ragaini}, {Rago}, {Rambaux}, {Ramos-Lerate},
  {Ranalli}, {Rauw}, {Read}, {Regibo}, {Renk}, {Reyl{\'e}}, {Ribeiro},
  {Rimoldini}, {Ripepi}, {Riva}, {Rixon}, {Roelens}, {Romero-G{\'o}mez},
  {Rowell}, {Royer}, {Rudolph}, {Ruiz-Dern}, {Sadowski}, {Sagrist{\`a}
  Sell{\'e}s}, {Sahlmann}, {Salgado}, {Salguero}, {Sarasso}, {Savietto},
  {Schnorhk}, {Schultheis}, {Sciacca}, {Segol}, {Segovia}, {Segransan},
  {Serpell}, {Shih}, {Smareglia}, {Smart}, {Smith}, {Solano}, {Solitro},
  {Sordo}, {Soria Nieto}, {Souchay}, {Spagna}, {Spoto}, {Stampa}, {Steele},
  {Steidelm{\"u}ller}, {Stephenson}, {Stoev}, {Suess}, {S{\"u}veges}, {Surdej},
  {Szabados}, {Szegedi-Elek}, {Tapiador}, {Taris}, {Tauran}, {Taylor},
  {Teixeira}, {Terrett}, {Tingley}, {Trager}, {Turon}, {Ulla}, {Utrilla},
  {Valentini}, {van Elteren}, {Van Hemelryck}, {van Leeuwen}, {Varadi},
  {Vecchiato}, {Veljanoski}, {Via}, {Vicente}, {Vogt}, {Voss}, {Votruba},
  {Voutsinas}, {Walmsley}, {Weiler}, {Weingrill}, {Werner}, {Wevers},
  {Whitehead}, {Wyrzykowski}, {Yoldas}, {{\v{Z}}erjal}, {Zucker}, {Zurbach},
  {Zwitter}, {Alecu}, {Allen}, {Allende Prieto}, {Amorim},
  {Anglada-Escud{\'e}}, {Arsenijevic}, {Azaz}, {Balm}, {Beck}, {Bernstein},
  {Bigot}, {Bijaoui}, {Blasco}, {Bonfigli}, {Bono}, {Boudreault}, {Bressan},
  {Brown}, {Brunet}, {Bunclark}, {Buonanno}, {Butkevich}, {Carret}, {Carrion},
  {Chemin}, {Ch{\'e}reau}, {Corcione}, {Darmigny}, {de Boer}, {de Teodoro}, {de
  Zeeuw}, {Delle Luche}, {Domingues}, {Dubath}, {Fodor}, {Fr{\'e}zouls},
  {Fries}, {Fustes}, {Fyfe}, {Gallardo}, {Gallegos}, {Gardiol}, {Gebran},
  {Gomboc}, {G{\'o}mez}, {Grux}, {Gueguen}, {Heyrovsky}, {Hoar}, {Iannicola},
  {Isasi Parache}, {Janotto}, {Joliet}, {Jonckheere}, {Keil}, {Kim},
  {Klagyivik}, {Klar}, {Knude}, {Kochukhov}, {Kolka}, {Kos}, {Kutka}, {Lainey},
  {LeBouquin}, {Liu}, {Loreggia}, {Makarov}, {Marseille}, {Martayan},
  {Martinez-Rubi}, {Massart}, {Meynadier}, {Mignot}, {Munari}, {Nguyen},
  {Nordlander}, {Ocvirk}, {O'Flaherty}, {Olias Sanz}, {Ortiz}, {Osorio},
  {Oszkiewicz}, {Ouzounis}, {Palmer}, {Park}, {Pasquato}, {Peltzer}, {Peralta},
  {P{\'e}turaud}, {Pieniluoma}, {Pigozzi}, {Poels}, {Prat}, {Prod'homme},
  {Raison}, {Rebordao}, {Risquez}, {Rocca-Volmerange}, {Rosen}, {Ruiz-Fuertes},
  {Russo}, {Sembay}, {Serraller Vizcaino}, {Short}, {Siebert}, {Silva},
  {Sinachopoulos}, {Slezak}, {Soffel}, {Sosnowska}, {Strai{\v{z}}ys}, {ter
  Linden}, {Terrell}, {Theil}, {Tiede}, {Troisi}, {Tsalmantza}, {Tur},
  {Vaccari}, {Vachier}, {Valles}, {Van Hamme}, {Veltz}, {Virtanen}, {Wallut},
  {Wichmann}, {Wilkinson}, {Ziaeepour}, \& {Zschocke}}]{GaiaDR2}
{Gaia Collaboration}, {Prusti}, T., {de Bruijne}, J.~H.~J., {et~al.} 2016,
  \aap, 595, A1, \dodoi{10.1051/0004-6361/201629272}

\bibitem[{{Gaia Collaboration} {et~al.}(2021){Gaia Collaboration}, {Brown},
  {Vallenari}, {Prusti}, {de Bruijne}, {Babusiaux}, {Biermann}, {Creevey},
  {Evans}, {Eyer}, {Hutton}, {Jansen}, {Jordi}, {Klioner}, {Lammers},
  {Lindegren}, {Luri}, {Mignard}, {Panem}, {Pourbaix}, {Randich}, {Sartoretti},
  {Soubiran}, {Walton}, {Arenou}, {Bailer-Jones}, {Bastian}, {Cropper},
  {Drimmel}, {Katz}, {Lattanzi}, {van Leeuwen}, {Bakker}, {Cacciari},
  {Casta{\~n}eda}, {De Angeli}, {Ducourant}, {Fabricius}, {Fouesneau},
  {Fr{\'e}mat}, {Guerra}, {Guerrier}, {Guiraud}, {Jean-Antoine Piccolo},
  {Masana}, {Messineo}, {Mowlavi}, {Nicolas}, {Nienartowicz}, {Pailler},
  {Panuzzo}, {Riclet}, {Roux}, {Seabroke}, {Sordo}, {Tanga}, {Th{\'e}venin},
  {Gracia-Abril}, {Portell}, {Teyssier}, {Altmann}, {Andrae}, {Bellas-Velidis},
  {Benson}, {Berthier}, {Blomme}, {Brugaletta}, {Burgess}, {Busso}, {Carry},
  {Cellino}, {Cheek}, {Clementini}, {Damerdji}, {Davidson}, {Delchambre},
  {Dell'Oro}, {Fern{\'a}ndez-Hern{\'a}ndez}, {Galluccio}, {Garc{\'\i}a-Lario},
  {Garcia-Reinaldos}, {Gonz{\'a}lez-N{\'u}{\~n}ez}, {Gosset}, {Haigron},
  {Halbwachs}, {Hambly}, {Harrison}, {Hatzidimitriou}, {Heiter},
  {Hern{\'a}ndez}, {Hestroffer}, {Hodgkin}, {Holl}, {Jan{\ss}en}, {Jevardat de
  Fombelle}, {Jordan}, {Krone-Martins}, {Lanzafame}, {L{\"o}ffler}, {Lorca},
  {Manteiga}, {Marchal}, {Marrese}, {Moitinho}, {Mora}, {Muinonen}, {Osborne},
  {Pancino}, {Pauwels}, {Petit}, {Recio-Blanco}, {Richards}, {Riello},
  {Rimoldini}, {Robin}, {Roegiers}, {Rybizki}, {Sarro}, {Siopis}, {Smith},
  {Sozzetti}, {Ulla}, {Utrilla}, {van Leeuwen}, {van Reeven}, {Abbas}, {Abreu
  Aramburu}, {Accart}, {Aerts}, {Aguado}, {Ajaj}, {Altavilla}, {{\'A}lvarez},
  {{\'A}lvarez Cid-Fuentes}, {Alves}, {Anderson}, {Anglada Varela}, {Antoja},
  {Audard}, {Baines}, {Baker}, {Balaguer-N{\'u}{\~n}ez}, {Balbinot}, {Balog},
  {Barache}, {Barbato}, {Barros}, {Barstow}, {Bartolom{\'e}}, {Bassilana},
  {Bauchet}, {Baudesson-Stella}, {Becciani}, {Bellazzini}, {Bernet}, {Bertone},
  {Bianchi}, {Blanco-Cuaresma}, {Boch}, {Bombrun}, {Bossini}, {Bouquillon},
  {Bragaglia}, {Bramante}, {Breedt}, {Bressan}, {Brouillet}, {Bucciarelli},
  {Burlacu}, {Busonero}, {Butkevich}, {Buzzi}, {Caffau}, {Cancelliere},
  {C{\'a}novas}, {Cantat-Gaudin}, {Carballo}, {Carlucci}, {Carnerero},
  {Carrasco}, {Casamiquela}, {Castellani}, {Castro-Ginard}, {Castro Sampol},
  {Chaoul}, {Charlot}, {Chemin}, {Chiavassa}, {Cioni}, {Comoretto}, {Cooper},
  {Cornez}, {Cowell}, {Crifo}, {Crosta}, {Crowley}, {Dafonte}, {Dapergolas},
  {David}, {David}, {de Laverny}, {De Luise}, {De March}, {De Ridder}, {de
  Souza}, {de Teodoro}, {de Torres}, {del Peloso}, {del Pozo}, {Delbo},
  {Delgado}, {Delgado}, {Delisle}, {Di Matteo}, {Diakite}, {Diener},
  {Distefano}, {Dolding}, {Eappachen}, {Edvardsson}, {Enke}, {Esquej}, {Fabre},
  {Fabrizio}, {Faigler}, {Fedorets}, {Fernique}, {Fienga}, {Figueras},
  {Fouron}, {Fragkoudi}, {Fraile}, {Franke}, {Gai}, {Garabato},
  {Garcia-Gutierrez}, {Garc{\'\i}a-Torres}, {Garofalo}, {Gavras}, {Gerlach},
  {Geyer}, {Giacobbe}, {Gilmore}, {Girona}, {Giuffrida}, {Gomel}, {Gomez},
  {Gonzalez-Santamaria}, {Gonz{\'a}lez-Vidal}, {Granvik},
  {Guti{\'e}rrez-S{\'a}nchez}, {Guy}, {Hauser}, {Haywood}, {Helmi}, {Hidalgo},
  {Hilger}, {H{\l}adczuk}, {Hobbs}, {Holland}, {Huckle}, {Jasniewicz},
  {Jonker}, {Juaristi Campillo}, {Julbe}, {Karbevska}, {Kervella}, {Khanna},
  {Kochoska}, {Kontizas}, {Kordopatis}, {Korn}, {Kostrzewa-Rutkowska},
  {Kruszy{\'n}ska}, {Lambert}, {Lanza}, {Lasne}, {Le Campion}, {Le Fustec},
  {Lebreton}, {Lebzelter}, {Leccia}, {Leclerc}, {Lecoeur-Taibi}, {Liao},
  {Licata}, {Lindstr{\o}m}, {Lister}, {Livanou}, {Lobel}, {Madrero Pardo},
  {Managau}, {Mann}, {Marchant}, {Marconi}, {Marcos Santos}, {Marinoni},
  {Marocco}, {Marshall}, {Martin Polo}, {Mart{\'\i}n-Fleitas}, {Masip},
  {Massari}, {Mastrobuono-Battisti}, {Mazeh}, {McMillan}, {Messina},
  {Michalik}, {Millar}, {Mints}, {Molina}, {Molinaro}, {Moln{\'a}r},
  {Montegriffo}, {Mor}, {Morbidelli}, {Morel}, {Morris}, {Mulone}, {Munoz},
  {Muraveva}, {Murphy}, {Musella}, {Noval}, {Ord{\'e}novic}, {Orr{\`u}},
  {Osinde}, {Pagani}, {Pagano}, {Palaversa}, {Palicio}, {Panahi}, {Pawlak},
  {Pe{\~n}alosa Esteller}, {Penttil{\"a}}, {Piersimoni}, {Pineau}, {Plachy},
  {Plum}, {Poggio}, {Poretti}, {Poujoulet}, {Pr{\v{s}}a}, {Pulone}, {Racero},
  {Ragaini}, {Rainer}, {Raiteri}, {Rambaux}, {Ramos}, {Ramos-Lerate}, {Re
  Fiorentin}, {Regibo}, {Reyl{\'e}}, {Ripepi}, {Riva}, {Rixon}, {Robichon},
  {Robin}, {Roelens}, {Rohrbasser}, {Romero-G{\'o}mez}, {Rowell}, {Royer},
  {Rybicki}, {Sadowski}, {Sagrist{\`a} Sell{\'e}s}, {Sahlmann}, {Salgado},
  {Salguero}, {Samaras}, {Sanchez Gimenez}, {Sanna}, {Santove{\~n}a},
  {Sarasso}, {Schultheis}, {Sciacca}, {Segol}, {Segovia}, {S{\'e}gransan},
  {Semeux}, {Shahaf}, {Siddiqui}, {Siebert}, {Siltala}, {Slezak}, {Smart},
  {Solano}, {Solitro}, {Souami}, {Souchay}, {Spagna}, {Spoto}, {Steele},
  {Steidelm{\"u}ller}, {Stephenson}, {S{\"u}veges}, {Szabados}, {Szegedi-Elek},
  {Taris}, {Tauran}, {Taylor}, {Teixeira}, {Thuillot}, {Tonello}, {Torra},
  {Torra}, {Turon}, {Unger}, {Vaillant}, {van Dillen}, {Vanel}, {Vecchiato},
  {Viala}, {Vicente}, {Voutsinas}, {Weiler}, {Wevers}, {Wyrzykowski}, {Yoldas},
  {Yvard}, {Zhao}, {Zorec}, {Zucker}, {Zurbach}, \& {Zwitter}}]{GaiaEDR3_2021}
{Gaia Collaboration}, {Brown}, A.~G.~A., {Vallenari}, A., {et~al.} 2021, \aap,
  649, A1, \dodoi{10.1051/0004-6361/202039657}

\bibitem[{{Gaia Collaboration} {et~al.}(2022){Gaia Collaboration}, {Vallenari},
  {Brown}, {Prusti}, {de Bruijne}, {Arenou}, {Babusiaux}, {Biermann},
  {Creevey}, {Ducourant}, {Evans}, {Eyer}, {Guerra}, {Hutton}, {Jordi},
  {Klioner}, {Lammers}, {Lindegren}, {Luri}, {Mignard}, {Panem}, {Pourbaix},
  {Randich}, {Sartoretti}, {Soubiran}, {Tanga}, {Walton}, {Bailer-Jones},
  {Bastian}, {Drimmel}, {Jansen}, {Katz}, {Lattanzi}, {van Leeuwen}, {Bakker},
  {Cacciari}, {Casta{\~n}eda}, {De Angeli}, {Fabricius}, {Fouesneau},
  {Fr{\'e}mat}, {Galluccio}, {Guerrier}, {Heiter}, {Masana}, {Messineo},
  {Mowlavi}, {Nicolas}, {Nienartowicz}, {Pailler}, {Panuzzo}, {Riclet}, {Roux},
  {Seabroke}, {Sordo{\o}rcit}, {Th{\'e}venin}, {Gracia-Abril}, {Portell},
  {Teyssier}, {Altmann}, {Andrae}, {Audard}, {Bellas-Velidis}, {Benson},
  {Berthier}, {Blomme}, {Burgess}, {Busonero}, {Busso}, {C{\'a}novas}, {Carry},
  {Cellino}, {Cheek}, {Clementini}, {Damerdji}, {Davidson}, {de Teodoro},
  {Nu{\~n}ez Campos}, {Delchambre}, {Dell'Oro}, {Esquej},
  {Fern{\'a}ndez-Hern{\'a}ndez}, {Fraile}, {Garabato}, {Garc{\'\i}a-Lario},
  {Gosset}, {Haigron}, {Halbwachs}, {Hambly}, {Harrison}, {Hern{\'a}ndez},
  {Hestroffer}, {Hodgkin}, {Holl}, {Jan{\ss}en}, {Jevardat de Fombelle},
  {Jordan}, {Krone-Martins}, {Lanzafame}, {L{\"o}ffler}, {Marchal}, {Marrese},
  {Moitinho}, {Muinonen}, {Osborne}, {Pancino}, {Pauwels}, {Recio-Blanco},
  {Reyl{\'e}}, {Riello}, {Rimoldini}, {Roegiers}, {Rybizki}, {Sarro}, {Siopis},
  {Smith}, {Sozzetti}, {Utrilla}, {van Leeuwen}, {Abbas}, {{\'A}brah{\'a}m},
  {Abreu Aramburu}, {Aerts}, {Aguado}, {Ajaj}, {Aldea-Montero}, {Altavilla},
  {{\'A}lvarez}, {Alves}, {Anders}, {Anderson}, {Anglada Varela}, {Antoja},
  {Baines}, {Baker}, {Balaguer-N{\'u}{\~n}ez}, {Balbinot}, {Balog}, {Barache},
  {Barbato}, {Barros}, {Barstow}, {Bartolom{\'e}}, {Bassilana}, {Bauchet},
  {Becciani}, {Bellazzini}, {Berihuete}, {Bernet}, {Bertone}, {Bianchi},
  {Binnenfeld}, {Blanco-Cuaresma}, {Blazere}, {Boch}, {Bombrun}, {Bossini},
  {Bouquillon}, {Bragaglia}, {Bramante}, {Breedt}, {Bressan}, {Brouillet},
  {Brugaletta}, {Bucciarelli}, {Burlacu}, {Butkevich}, {Buzzi}, {Caffau},
  {Cancelliere}, {Cantat-Gaudin}, {Carballo}, {Carlucci}, {Carnerero},
  {Carrasco}, {Casamiquela}, {Castellani}, {Castro-Ginard}, {Chaoul},
  {Charlot}, {Chemin}, {Chiaramida}, {Chiavassa}, {Chornay}, {Comoretto},
  {Contursi}, {Cooper}, {Cornez}, {Cowell}, {Crifo}, {Cropper}, {Crosta},
  {Crowley}, {Dafonte}, {Dapergolas}, {David}, {David}, {de Laverny}, {De
  Luise}, {De March}, {De Ridder}, {de Souza}, {de Torres}, {del Peloso}, {del
  Pozo}, {Delbo}, {Delgado}, {Delisle}, {Demouchy}, {Dharmawardena}, {Di
  Matteo}, {Diakite}, {Diener}, {Distefano}, {Dolding}, {Edvardsson}, {Enke},
  {Fabre}, {Fabrizio}, {Faigler}, {Fedorets}, {Fernique}, {Fienga}, {Figueras},
  {Fournier}, {Fouron}, {Fragkoudi}, {Gai}, {Garcia-Gutierrez},
  {Garcia-Reinaldos}, {Garc{\'\i}a-Torres}, {Garofalo}, {Gavel}, {Gavras},
  {Gerlach}, {Geyer}, {Giacobbe}, {Gilmore}, {Girona}, {Giuffrida}, {Gomel},
  {Gomez}, {Gonz{\'a}lez-N{\'u}{\~n}ez}, {Gonz{\'a}lez-Santamar{\'\i}a},
  {Gonz{\'a}lez-Vidal}, {Granvik}, {Guillout}, {Guiraud},
  {Guti{\'e}rrez-S{\'a}nchez}, {Guy}, {Hatzidimitriou}, {Hauser}, {Haywood},
  {Helmer}, {Helmi}, {Sarmiento}, {Hidalgo}, {Hilger}, {H{\l}adczuk}, {Hobbs},
  {Holland}, {Huckle}, {Jardine}, {Jasniewicz}, {Jean-Antoine Piccolo},
  {Jim{\'e}nez-Arranz}, {Jorissen}, {Juaristi Campillo}, {Julbe}, {Karbevska},
  {Kervella}, {Khanna}, {Kontizas}, {Kordopatis}, {Korn}, {K{\'o}sp{\'a}l},
  {Kostrzewa-Rutkowska}, {Kruszy{\'n}ska}, {Kun}, {Laizeau}, {Lambert},
  {Lanza}, {Lasne}, {Le Campion}, {Lebreton}, {Lebzelter}, {Leccia}, {Leclerc},
  {Lecoeur-Taibi}, {Liao}, {Licata}, {Lindstr{\o}m}, {Lister}, {Livanou},
  {Lobel}, {Lorca}, {Loup}, {Madrero Pardo}, {Magdaleno Romeo}, {Managau},
  {Mann}, {Manteiga}, {Marchant}, {Marconi}, {Marcos}, {Marcos Santos},
  {Mar{\'\i}n Pina}, {Marinoni}, {Marocco}, {Marshall}, {Polo},
  {Mart{\'\i}n-Fleitas}, {Marton}, {Mary}, {Masip}, {Massari},
  {Mastrobuono-Battisti}, {Mazeh}, {McMillan}, {Messina}, {Michalik}, {Millar},
  {Mints}, {Molina}, {Molinaro}, {Moln{\'a}r}, {Monari}, {Mongui{\'o}},
  {Montegriffo}, {Montero}, {Mor}, {Mora}, {Morbidelli}, {Morel}, {Morris},
  {Muraveva}, {Murphy}, {Musella}, {Nagy}, {Noval}, {Oca{\~n}a}, {Ogden},
  {Ordenovic}, {Osinde}, {Pagani}, {Pagano}, {Palaversa}, {Palicio},
  {Pallas-Quintela}, {Panahi}, {Payne-Wardenaar}, {Pe{\~n}alosa Esteller},
  {Penttil{\"a}}, {Pichon}, {Piersimoni}, {Pineau}, {Plachy}, {Plum}, {Poggio},
  {Pr{\v{s}}a}, {Pulone}, {Racero}, {Ragaini}, {Rainer}, {Raiteri}, {Rambaux},
  {Ramos}, {Ramos-Lerate}, {Re Fiorentin}, {Regibo}, {Richards}, {Rios Diaz},
  {Ripepi}, {Riva}, {Rix}, {Rixon}, {Robichon}, {Robin}, {Robin}, {Roelens},
  {Rogues}, {Rohrbasser}, {Romero-G{\'o}mez}, {Rowell}, {Royer}, {Ruz Mieres},
  {Rybicki}, {Sadowski}, {S{\'a}ez N{\'u}{\~n}ez}, {Sagrist{\`a} Sell{\'e}s},
  {Sahlmann}, {Salguero}, {Samaras}, {Sanchez Gimenez}, {Sanna},
  {Santove{\~n}a}, {Sarasso}, {Schultheis}, {Sciacca}, {Segol}, {Segovia},
  {S{\'e}gransan}, {Semeux}, {Shahaf}, {Siddiqui}, {Siebert}, {Siltala},
  {Silvelo}, {Slezak}, {Slezak}, {Smart}, {Snaith}, {Solano}, {Solitro},
  {Souami}, {Souchay}, {Spagna}, {Spina}, {Spoto}, {Steele},
  {Steidelm{\"u}ller}, {Stephenson}, {S{\"u}veges}, {Surdej}, {Szabados},
  {Szegedi-Elek}, {Taris}, {Taylo}, {Teixeira}, {Tolomei}, {Tonello}, {Torra},
  {Torra}, {Torralba Elipe}, {Trabucchi}, {Tsounis}, {Turon}, {Ulla}, {Unger},
  {Vaillant}, {van Dillen}, {van Reeven}, {Vanel}, {Vecchiato}, {Viala},
  {Vicente}, {Voutsinas}, {Weiler}, {Wevers}, {Wyrzykowski}, {Yoldas}, {Yvard},
  {Zhao}, {Zorec}, {Zucker}, \& {Zwitter}}]{GaiaDR3_summary}
{Gaia Collaboration}, {Vallenari}, A., {Brown}, A.~G.~A., {et~al.} 2022, arXiv
  e-prints, arXiv:2208.00211.
\newblock \doarXiv{2208.00211}

\bibitem[{{Green}(2019)}]{SNRcata_Green19}
{Green}, D.~A. 2019, Journal of Astrophysics and Astronomy, 40, 36,
  \dodoi{10.1007/s12036-019-9601-6}

\bibitem[{{Green}(2018)}]{dustmaps}
{Green}, G. 2018, The Journal of Open Source Software, 3, 695,
  \dodoi{10.21105/joss.00695}

\bibitem[{{Green} {et~al.}(2019){Green}, {Schlafly}, {Zucker}, {Speagle}, \&
  {Finkbeiner}}]{Green19}
{Green}, G.~M., {Schlafly}, E., {Zucker}, C., {Speagle}, J.~S., \&
  {Finkbeiner}, D. 2019, \apj, 887, 93, \dodoi{10.3847/1538-4357/ab5362}

\bibitem[{{Green} {et~al.}(2015){Green}, {Schlafly}, {Finkbeiner}, {Rix},
  {Martin}, {Burgett}, {Draper}, {Flewelling}, {Hodapp}, {Kaiser}, {Kudritzki},
  {Magnier}, {Metcalfe}, {Price}, {Tonry}, \& {Wainscoat}}]{green15}
{Green}, G.~M., {Schlafly}, E.~F., {Finkbeiner}, D.~P., {et~al.} 2015, \apj,
  810, 25, \dodoi{10.1088/0004-637X/810/1/25}

\bibitem[{{Gro{\ss}schedl} {et~al.}(2018){Gro{\ss}schedl}, {Alves}, {Meingast},
  {Ackerl}, {Ascenso}, {Bouy}, {Burkert}, {Forbrich}, {F{\"u}rnkranz},
  {Goodman}, {Hacar}, {Herbst-Kiss}, {Lada}, {Larreina}, {Leschinski},
  {Lombardi}, {Moitinho}, {Mortimer}, \& {Zari}}]{OrionA_YSO_18}
{Gro{\ss}schedl}, J.~E., {Alves}, J., {Meingast}, S., {et~al.} 2018, \aap, 619,
  A106, \dodoi{10.1051/0004-6361/201833901}

\bibitem[{{Jian} {et~al.}(2017){Jian}, {Gao}, {Zhao}, \&
  {Jiang}}]{Jian2017_RevBE}
{Jian}, M., {Gao}, S., {Zhao}, H., \& {Jiang}, B. 2017, \aj, 153, 5,
  \dodoi{10.3847/1538-3881/153/1/5}

\bibitem[{{Jordi} {et~al.}(2010){Jordi}, {Gebran}, {Carrasco}, {de Bruijne},
  {Voss}, {Fabricius}, {Knude}, {Vallenari}, {Kohley}, \&
  {Mora}}]{Jordi_2010_band}
{Jordi}, C., {Gebran}, M., {Carrasco}, J.~M., {et~al.} 2010, \aap, 523, A48,
  \dodoi{10.1051/0004-6361/201015441}

\bibitem[{{Leike} {et~al.}(2020){Leike}, {Glatzle}, \& {En{\ss}lin}}]{L20}
{Leike}, R.~H., {Glatzle}, M., \& {En{\ss}lin}, T.~A. 2020, \aap, 639, A138,
  \dodoi{10.1051/0004-6361/202038169}

\bibitem[{{Lombardi} {et~al.}(2011){Lombardi}, {Alves}, \&
  {Lada}}]{Orion2011IV}
{Lombardi}, M., {Alves}, J., \& {Lada}, C.~J. 2011, \aap, 535, A16,
  \dodoi{10.1051/0004-6361/201116915}

\bibitem[{{Lombardi} {et~al.}(2010){Lombardi}, {Lada}, \&
  {Alves}}]{TauCaliPer2010III}
{Lombardi}, M., {Lada}, C.~J., \& {Alves}, J. 2010, \aap, 512, A67,
  \dodoi{10.1051/0004-6361/200912670}

\bibitem[{{Luo} {et~al.}(2015){Luo}, {Zhao}, {Zhao}, {Deng}, {Liu}, {Jing},
  {Wang}, {Zhang}, {Shi}, {Cui}, {Chu}, {Li}, {Bai}, {Wu}, {Cai}, {Cao}, {Cao},
  {Carlin}, {Chen}, {Chen}, {Chen}, {Chen}, {Chen}, {Chen}, {Chen},
  {Christlieb}, {Chu}, {Cui}, {Dong}, {Du}, {Fan}, {Feng}, {Fu}, {Gao}, {Gong},
  {Gu}, {Guo}, {Han}, {He}, {Hou}, {Hou}, {Hou}, {Hu}, {Hu}, {Hu}, {Huo},
  {Jia}, {Jiang}, {Jiang}, {Jiang}, {Jin}, {Kong}, {Kong}, {Lei}, {Li}, {Li},
  {Li}, {Li}, {Li}, {Li}, {Li}, {Li}, {Li}, {Li}, {Li}, {Li}, {Liang}, {Lin},
  {Liu}, {Liu}, {Liu}, {Liu}, {Lu}, {Luo}, {Mao}, {Newberg}, {Ni}, {Qi}, {Qi},
  {Shen}, {Shi}, {Song}, {Song}, {Su}, {Su}, {Tang}, {Tao}, {Tian}, {Wang},
  {Wang}, {Wang}, {Wang}, {Wang}, {Wang}, {Wang}, {Wang}, {Wang}, {Wang},
  {Wang}, {Wang}, {Wang}, {Wang}, {Wang}, {Wang}, {Wang}, {Wang}, {Wang},
  {Wang}, {Wei}, {Wei}, {Wu}, {Wu}, {Wu}, {Wu}, {Xing}, {Xu}, {Xu}, {Xu},
  {Yan}, {Yang}, {Yang}, {Yang}, {Yang}, {Yao}, {Yu}, {Yuan}, {Yuan}, {Yuan},
  {Yuan}, {Zhai}, {Zhang}, {Zhang}, {Zhang}, {Zhang}, {Zhang}, {Zhang},
  {Zhang}, {Zhang}, {Zhao}, {Zhou}, {Zhou}, {Zhu}, {Zhu}, {Zou}, \&
  {Zuo}}]{Luo15_LMT}
{Luo}, A.~L., {Zhao}, Y.-H., {Zhao}, G., {et~al.} 2015, Research in Astronomy
  and Astrophysics, 15, 1095, \dodoi{10.1088/1674-4527/15/8/002}

\bibitem[{{Magnani} {et~al.}(1985){Magnani}, {Blitz}, \& {Mundy}}]{MBMlist85}
{Magnani}, L., {Blitz}, L., \& {Mundy}, L. 1985, \apj, 295, 402,
  \dodoi{10.1086/163385}

\bibitem[{{Majewski} {et~al.}(2017){Majewski}, {Schiavon}, {Frinchaboy},
  {Allende Prieto}, {Barkhouser}, {Bizyaev}, {Blank}, {Brunner}, {Burton},
  {Carrera}, {Chojnowski}, {Cunha}, {Epstein}, {Fitzgerald}, {Garc{\'\i}a
  P{\'e}rez}, {Hearty}, {Henderson}, {Holtzman}, {Johnson}, {Lam}, {Lawler},
  {Maseman}, {M{\'e}sz{\'a}ros}, {Nelson}, {Nguyen}, {Nidever}, {Pinsonneault},
  {Shetrone}, {Smee}, {Smith}, {Stolberg}, {Skrutskie}, {Walker}, {Wilson},
  {Zasowski}, {Anders}, {Basu}, {Beland}, {Blanton}, {Bovy}, {Brownstein},
  {Carlberg}, {Chaplin}, {Chiappini}, {Eisenstein}, {Elsworth}, {Feuillet},
  {Fleming}, {Galbraith-Frew}, {Garc{\'\i}a}, {Garc{\'\i}a-Hern{\'a}ndez},
  {Gillespie}, {Girardi}, {Gunn}, {Hasselquist}, {Hayden}, {Hekker}, {Ivans},
  {Kinemuchi}, {Klaene}, {Mahadevan}, {Mathur}, {Mosser}, {Muna}, {Munn},
  {Nichol}, {O'Connell}, {Parejko}, {Robin}, {Rocha-Pinto}, {Schultheis},
  {Serenelli}, {Shane}, {Silva Aguirre}, {Sobeck}, {Thompson}, {Troup},
  {Weinberg}, \& {Zamora}}]{APOGEE_review}
{Majewski}, S.~R., {Schiavon}, R.~P., {Frinchaboy}, P.~M., {et~al.} 2017, \aj,
  154, 94, \dodoi{10.3847/1538-3881/aa784d}

\bibitem[{{Ortiz-Le{\'o}n} {et~al.}(2018){Ortiz-Le{\'o}n}, {Loinard}, {Dzib},
  {Galli}, {Kounkel}, {Mioduszewski}, {Rodr{\'\i}guez}, {Torres}, {Hartmann},
  {Boden}, {Evans}, {Brice{\~n}o}, \& {Tobin}}]{Per_Go_YSO18}
{Ortiz-Le{\'o}n}, G.~N., {Loinard}, L., {Dzib}, S.~A., {et~al.} 2018, \apj,
  865, 73, \dodoi{10.3847/1538-4357/aada49}

\bibitem[{Pedregosa {et~al.}(2011)Pedregosa, Varoquaux, Gramfort, Michel,
  Thirion, Grisel, Blondel, Prettenhofer, Weiss, Dubourg, Vanderplas, Passos,
  Cournapeau, Brucher, Perrot, \& Duchesnay}]{scikit-learn}
Pedregosa, F., Varoquaux, G., Gramfort, A., {et~al.} 2011, Journal of Machine
  Learning Research, 12, 2825

\bibitem[{{Rezaei Kh.} {et~al.}(2020){Rezaei Kh.}, {Bailer-Jones}, {Soler}, \&
  {Zari}}]{Orion2022_Kh}
{Rezaei Kh.}, S., {Bailer-Jones}, C. A.~L., {Soler}, J.~D., \& {Zari}, E. 2020,
  \aap, 643, A151, \dodoi{10.1051/0004-6361/202038708}

\bibitem[{{Rezaei Kh.} \& {Kainulainen}(2022)}]{CaliOriA21_Kh}
{Rezaei Kh.}, S., \& {Kainulainen}, J. 2022, \apjl, 930, L22,
  \dodoi{10.3847/2041-8213/ac67db}

\bibitem[{{Schlafly} {et~al.}(2014){Schlafly}, {Green}, {Finkbeiner}, {Rix},
  {Bell}, {Burgett}, {Chambers}, {Draper}, {Hodapp}, {Kaiser}, {Magnier},
  {Martin}, {Metcalfe}, {Price}, \& {Tonry}}]{Sch14_MCsDis}
{Schlafly}, E.~F., {Green}, G., {Finkbeiner}, D.~P., {et~al.} 2014, \apj, 786,
  29, \dodoi{10.1088/0004-637X/786/1/29}

\bibitem[{{Schlegel} {et~al.}(1998){Schlegel}, {Finkbeiner}, \&
  {Davis}}]{sfd98}
{Schlegel}, D.~J., {Finkbeiner}, D.~P., \& {Davis}, M. 1998, \apj, 500, 525,
  \dodoi{10.1086/305772}

\bibitem[{{Skrutskie} {et~al.}(2006){Skrutskie}, {Cutri}, {Stiening},
  {Weinberg}, {Schneider}, {Carpenter}, {Beichman}, {Capps}, {Chester},
  {Elias}, {Huchra}, {Liebert}, {Lonsdale}, {Monet}, {Price}, {Seitzer},
  {Jarrett}, {Kirkpatrick}, {Gizis}, {Howard}, {Evans}, {Fowler}, {Fullmer},
  {Hurt}, {Light}, {Kopan}, {Marsh}, {McCallon}, {Tam}, {Van Dyk}, \&
  {Wheelock}}]{2MASS_2006}
{Skrutskie}, M.~F., {Cutri}, R.~M., {Stiening}, R., {et~al.} 2006, \aj, 131,
  1163, \dodoi{10.1086/498708}

\bibitem[{{Sun} {et~al.}(2021{\natexlab{a}}){Sun}, {Jiang}, {Yuan}, \&
  {Li}}]{Sun21_G_BV}
{Sun}, M., {Jiang}, B., {Yuan}, H., \& {Li}, J. 2021{\natexlab{a}}, \apjs, 254,
  38, \dodoi{10.3847/1538-4365/abf929}

\bibitem[{{Sun} {et~al.}(2021{\natexlab{b}}){Sun}, {Jiang}, {Zhao}, \&
  {Ren}}]{Sun21_MCdistance}
{Sun}, M., {Jiang}, B., {Zhao}, H., \& {Ren}, Y. 2021{\natexlab{b}}, \apjs,
  256, 46, \dodoi{10.3847/1538-4365/ac1601}

\bibitem[{{Sun} {et~al.}(2018){Sun}, {Jiang}, {Zhao}, {Gao}, {Gao}, {Jian}, \&
  {Yuan}}]{Sun18_BE}
{Sun}, M., {Jiang}, B.~W., {Zhao}, H., {et~al.} 2018, \apj, 861, 153,
  \dodoi{10.3847/1538-4357/aac776}

\bibitem[{{Vehtari} {et~al.}(2019){Vehtari}, {Gelman}, {Simpson}, {Carpenter},
  \& {B{\"u}rkner}}]{R_hat_19}
{Vehtari}, A., {Gelman}, A., {Simpson}, D., {Carpenter}, B., \& {B{\"u}rkner},
  P.-C. 2019, arXiv e-prints, arXiv:1903.08008.
\newblock \doarXiv{1903.08008}

\bibitem[{{Wang} \& {Chen}(2019)}]{Wang19_law}
{Wang}, S., \& {Chen}, X. 2019, \apj, 877, 116,
  \dodoi{10.3847/1538-4357/ab1c61}

\bibitem[{{Wang} \& {Jiang}(2014)}]{W2014_IRlaw}
{Wang}, S., \& {Jiang}, B.~W. 2014, \apjl, 788, L12,
  \dodoi{10.1088/2041-8205/788/1/L12}

\bibitem[{{Xue} {et~al.}(2016){Xue}, {Jiang}, {Gao}, {Liu}, {Wang}, \&
  {Li}}]{Xue2016_IRlaw}
{Xue}, M., {Jiang}, B.~W., {Gao}, J., {et~al.} 2016, \apjs, 224, 23,
  \dodoi{10.3847/0067-0049/224/2/23}

\bibitem[{{Yan} {et~al.}(2019{\natexlab{a}}){Yan}, {Zhang}, {Xu}, {Guo},
  {Macquart}, {Tang}, \& {Walsh}}]{Tau_zitai}
{Yan}, Q.-Z., {Zhang}, B., {Xu}, Y., {et~al.} 2019{\natexlab{a}}, \aap, 624,
  A6, \dodoi{10.1051/0004-6361/201834337}

\bibitem[{{Yan} {et~al.}(2019{\natexlab{b}}){Yan}, {Zhang}, {Xu}, {Guo},
  {Macquart}, {Tang}, \& {Walsh}}]{Yan19_MCsdis}
---. 2019{\natexlab{b}}, \aap, 624, A6, \dodoi{10.1051/0004-6361/201834337}

\bibitem[{{Zhao} {et~al.}(2018){Zhao}, {Jiang}, {Gao}, {Li}, \& {Sun}}]{Zhao18}
{Zhao}, H., {Jiang}, B., {Gao}, S., {Li}, J., \& {Sun}, M. 2018, \apj, 855, 12,
  \dodoi{10.3847/1538-4357/aaacd0}

\bibitem[{{Zhao} {et~al.}(2020){Zhao}, {Jiang}, {Li}, {Chen}, {Yu}, \&
  {Wang}}]{Zhao20}
{Zhao}, H., {Jiang}, B., {Li}, J., {et~al.} 2020, \apj, 891, 137,
  \dodoi{10.3847/1538-4357/ab75ef}

\bibitem[{{Zucker} {et~al.}(2018){Zucker}, {Schlafly}, {Speagle}, {Green},
  {Portillo}, {Finkbeiner}, \& {Goodman}}]{Per_zucker18}
{Zucker}, C., {Schlafly}, E.~F., {Speagle}, J.~S., {et~al.} 2018, \apj, 869,
  83, \dodoi{10.3847/1538-4357/aae97c}

\bibitem[{{Zucker} {et~al.}(2020){Zucker}, {Speagle}, {Schlafly}, {Green},
  {Finkbeiner}, {Goodman}, \& {Alves}}]{Zucker_20}
{Zucker}, C., {Speagle}, J.~S., {Schlafly}, E.~F., {et~al.} 2020, \aap, 633,
  A51, \dodoi{10.1051/0004-6361/201936145}

\bibitem[{{Zucker} {et~al.}(2019){Zucker}, {Speagle}, {Schlafly}, {Green},
  {Finkbeiner}, {Goodman}, \& {Alves}}]{Zucker19_MBM}
---. 2019, \apj, 879, 125, \dodoi{10.3847/1538-4357/ab2388}

\bibitem[{{Zucker} {et~al.}(2021){Zucker}, {Goodman}, {Alves}, {Bialy}, {Koch},
  {Speagle}, {Foley}, {Finkbeiner}, {Leike}, {En{\ss}lin}, {Peek}, \&
  {Edenhofer}}]{localMC21}
{Zucker}, C., {Goodman}, A., {Alves}, J., {et~al.} 2021, \apj, 919, 35,
  \dodoi{10.3847/1538-4357/ac1f96}

\end{thebibliography}


\begin{table}
\caption{The region and number of stars in the four MCs sightline}\label{Cloudboundary}
\centering
\begin{tabularx}{\textwidth}{XXXp{2.5cm}Xp{4cm}}
\hline\hline
    & \multicolumn{3}{c}{Coordinate} & \multicolumn{2}{c}{Number of Star}  \\
    \cmidrule(lr){2-4}\cmidrule(lr){5-6}
    & $l$ (\degr) & $b$ (\degr) & distance (pc) & until 600pc & within the distance range \\
\hline
Taurus & 155$\sim$195 & -40$\sim$10 & 100$\sim$250 & $\sim$140000 & $\sim$18000 \\
Orion & 180$\sim$220 & -25$\sim$0 & 350$\sim$500 & $\sim$47000 &$\sim$19000 \\
Perseus & 150$\sim$170 & -40$\sim$0 & 250$\sim$350 & $\sim$57000 &$\sim$11000 \\
California & 150$\sim$180 & -15$\sim$0 & 400$\sim$600 & $\sim$33000 & $\sim$18000 \\
\hline
\end{tabularx}
\end{table}

\begin{table*}
\caption{Summary of the substructures identified in the four molecular clouds }\label{MCsTable}
\label{DistanceTable}
\centering
\begin{threeparttable}
\begin{tabularx}{\textwidth}{p{2.3cm}p{1.4cm}p{1.4cm}p{1.3cm}p{1.2cm}p{1.2cm}p{1.2cm}p{1.2cm}p{4cm}}
\hline\hline
& \multicolumn{2}{c}{Coordinate}  & \multicolumn{3}{c}{Distance (pc)}  & \multicolumn{2}{c}{Max Color Excess (mag)} &  \\
\cmidrule(lr){2-3}\cmidrule(lr){4-6}\cmidrule(lr){7-8}
Component        &$l$ (\degr)  &$b$ (\degr)  & this work  & Z+21 & D+22  & $\BpRp{}$  &$\JK{}$ &Notes\tnote{(1)} \\
\hline
\textbf{Taurus}: &  &  &  &  &  &  &  & 135$\pm$20pc (S+14);\\
                 &  &  &  &  &  &  &  & $145^{+12}_{-16}$pc (Y+19)\\
TMC1             &($176,165,$ $166,177$)\tnote{(2)} &($-11,-15,$ $-18,-14$)\tnote{(2)} &129$\sim$157 &{131$\sim$168} & 145$\sim$190 &1.52 &0.68 & \\
TMC2             &($176,168,$ $169,177$)\tnote{(2)} &($-14,-18,$ $-21,-17$)\tnote{(2)} &132$\sim$156 &{131$\sim$168} & 110$\sim$145 &1.05 &0.50 &\\
TMC Filament     &($174,171,$ $161,169$)\tnote{(2)} &($-10,-12,$ $-6,-3$)\tnote{(2)}   &near 174   &  &  &0.99  &0.41  & \\
                 &  &  &  &  &  &  &  &  \\

\textbf{Perseus}:&  &  &  &  &  &  &  &260$\pm$26pc$\sim$315$\pm$32pc (S+14);\\
                 &  &  &  &  &  &  &  &$310^{+4}_{-4}$pc (Y+19)\\
Perseus Main     & 157$\sim$162  & $-$23$\sim$$-$16  & 285$\sim$306  & 279$\sim$301  & 300$\sim$350  &1.02  &0.44  &\\
Per Arm1         &       & $-$35$\sim$$-$25 & 267$\sim$300  &  &  &0.81  &0.19  &\\
Per Arm2         &$<$159 & $-$17$\sim$$-$12 & near 300    &  &  &0.66  &0.26  &250$\sim$350pc(California Filament, D+22)\\
IC348            & 160.5    & $-$18.27  &304 & & 325 &0.88\tnote{(4)}  &0.36\tnote{(4)}  & 295$\pm$4pc (Z+18);  \\
                 &   &   &  &  &  &   &   & 321$\pm$27pc (O+18)  \\
NGC1333          & 158.34   & $-$20.64  &287 & & 325 &0.79\tnote{(4)}  &0.28\tnote{(4)}  & 299$\pm$3pc (Z+18);  \\
                 &   &   &  &  &  &   &   & 294$\pm$28pc (O+18)  \\
                 &  &  &  &  &  &  &  &\\

\textbf{Orion}: &  &  &  &  &  &  &  & 420$\pm$42pc (S+14); \\
                 &  &  &  &  &  &  &  & 321$\pm$27pc (Y+19)  \\
$\lambda$ Orionis& 195          & $-$12               & 380$\sim$400  & 375$\sim$397  & 360$\sim$470 &0.99 &0.40  & $405^{+4}_{-4}$pc (Y+19); $R=5\degr$\tnote{(3)}\\
Orion A          & 206$\sim$216 & $-$22$\sim$$-$17.5  & $\approx$ 420 & 391$\sim$445  & 380$\sim$430 &1.13 &0.50  &  393$\pm$25pc (G+18)\\
Orion B          & 202$\sim$209 & $-$18$\sim$$-$7     & $\approx$ 410  & 397$\sim$406  & 380$\sim$420 &1.67 &0.58  &\\
                 &  &  &  &  &  &  &  &  \\

\textbf{California}:&  &  &  &  &  &  &  &  410$\pm$41pc (S+14); \\
                    &  &  &  &  &  &  &  &  $500^{+7}_{-7}$pc (Y+19)\\
California Main: & 155$\sim$170 & $-$13$\sim$$-$5  & 440$\sim$504  &  &  &1.22 &0.42  &  \\
Bubble           & 155$\sim$160 & $-$13$\sim$$-$7  & 440$\sim$448  &  &  &0.86 &0.37 & clear in 455pc slice in R+22\\
Sheet            & 160$\sim$170 & $-$10$\sim$$-$5  & 448$\sim$504  &  &  &1.22 &0.42 & clear from 495pc to 515pc slice in R+22\\
                 &  &  &  &  &  &  \\

\textbf{Others}:          &  &  &  &  &  &  &  &\\
Tau Per Shell    & 161.1     & $-$22.7   & 218            &  &  &  &  & Shell center is at 218pc,\\
                 &           &           &                &  &  &  &  & $R = 21\degr.5$ at 218pc\tnote{(3)}.\\
Bow Structure    & 150       & $-$15     & 170$\sim$220   &  &  &  &  & $R = 30\degr$ at about 200pc\tnote{(3)}.\\
Tau Ring         & 179.5     & $-$14.2   & 147$\sim$214   &  &  &  &  & Ring center is at 179pc\tnote{(3)}. \\
R1               & 152       & $-$1      &                &  &  &$\sim$0.2\tnote{(4)}  &  & $R=13\degr$ at around 300pc\tnote{(3)}.\\
R2               & 150       & $-$5      &                &  &  &$\sim$0.2\tnote{(4)}  &  & $R=8\degr$ at around 300pc\tnote{(3)}.\\
R3               & 168       & $-$17     &                &  &  &$\sim$0.2\tnote{(4)}  &  & $R=10\degr$ at around 300pc\tnote{(3)}.\\
\hline
\end{tabularx}
\begin{tablenotes}
\footnotesize
\item[(1)]: Z+21: \citet{localMC21}; D+22: \citet{OCTP4cloud}; R+22: \citet{CaliOriA21_Kh}; G+18: \citet{OrionA_YSO_18}; Z+18: \citet{Per_zucker18}; O+18: \citet{Per_Go_YSO18}; S+14: \citet{Sch14_MCsDis}; Y+19: \citet{Yan19_MCsdis}. Note that the results of Z+21, D+22 and R+22 are from extinction or dust map of MCs, G+18 studied the complete Orion A but using stars as tracers, Z+18, O+18, S+14 and Y+19  derived from several sightlines within the MCs region other than the whole MCs.
\item[(2)]: The coordinate of irregular polygon region vertex is given in array by anti-clockwise direction.
\item[(3)]: The radius of the projection of the ring, bow or shell structure.
\item[(4)]: Column "Max Color Excess" means the maximum color excess derived from the extinction map in each component except for IC348 and NGC1333 and R1, R2 and R3. Indeed, 0.88mag in $\BpRp{}$ and 0.36mag in $\JK{}$ are the extinction of IC348, the same as for NGC1333, and 0.2mag in $\BpRp{}$ is the typical extinction of R1, R2 and R3.
\end{tablenotes}
\end{threeparttable}
\end{table*}

\clearpage

\clearpage
\end{CJK*}
\end{document}